\documentclass[a4paper,fleqn]{cas-sc}

\usepackage[numbers]{natbib}

\usepackage{booktabs}   
\usepackage{amssymb}    
\usepackage{pifont}     
\usepackage{array}
\usepackage{subfigure}

\makeatletter
\@fleqntrue
\makeatother

\usepackage{etoolbox}
\AtBeginEnvironment{keywords}{%
    \vspace{-10pt} 
    \setlength{\parskip}{2pt} 
}
\AtEndEnvironment{keywords}{%
    \vspace{-5pt} 
}

\usepackage{caption}

\usepackage{graphicx}

\usepackage[switch]{lineno}

\usepackage{hyperref}
\hypersetup{
    linkcolor=blue,    
    citecolor=blue,   
    urlcolor=blue,      
    pdfborder={0 0 0}, 
    linktoc=all,
}

\def\tsc#1{\csdef{#1}{\textsc{\lowercase{#1}}\xspace}}
\tsc{WGM}
\tsc{QE}
\tsc{EP}
\tsc{PMS}
\tsc{BEC}
\tsc{DE}

\begin{document}
\sloppy
\let\WriteBookmarks\relax
\def\floatpagepagefraction{1}
\def\textpagefraction{.001}
\shorttitle{Multi-Scale Attention UNet3+  (MSA-UNet3+)}
\shortauthors{Rayan Merghani Ahmed et~al.}

\title [mode = title]{MSA-UNet3+: Multi-Scale Attention UNet3+ with New Supervised Prototypical Contrastive Loss for Coronary DSA Image Segmentation}                        



                        


\affiliation[1]{organization={Shenzhen Institutes of Advanced Technology (SIAT)},
                addressline={Chinese Academy of Sciences (CAS)}, 
                city={Shenzhen},
                postcode={518055}, 
                country={China}}
\affiliation[2]{organization={University of Chinese Academy of Sciences (UCAS)},
                city={Beijing},
                postcode={101408}, 
                country={China}}
\affiliation[3]{organization={Department of Biomedical Engineering and Systems, Faculty of Engineering},
                addressline={Cairo University},
                city={Cairo},
                postcode={12613}, 
                country={Egypt}}

\affiliation[4]{organization={The Neurosurgery Department},
                addressline={General Hospital of Southern Theater Command}, 
                city={Guangzhou},
                postcode={510000}, 
                country={China}}
                
\affiliation[5]{organization={The Department of Biomedical Engineering},
                addressline={Air Force Hospital of Southern Theater Command of PLA},
                }


\author[1,2]{Rayan Merghani Ahmed}
\ead{rayan@siat.ac.cn }
\author[1,2]{Adnan Iltaf}
\ead{adnan@siat.ac.cn}

\author[3]{Mohamed Elmanna}
\ead{Mohamed_elmanna@hotmail.com}
\author[4]{Gang Zhao}
\ead{1443364859@qq.com}
\author[5]{Hongliang Li}
\ead{david_lhl@163.com}
\author[1]{Yue Du}
\ead{yue.du2@siat.ac.cn}

\author[1]{Bin Li}[orcid= 0000-0002-6508-5071]
\cormark[1]
\ead{b.li2@siat.ac.cn}
\author[1]{Shoujun Zhou}[orcid= 0000-0003-3232-6796]
\cormark[1]
\ead{sj.zhou@siat.ac.cn}







\cortext[cor1]{Corresponding author}


\begin{abstract}
Accurate segmentation of coronary Digital Subtraction Angiography (DSA) images is essential to diagnose and treat coronary artery diseases (CAD). Despite advances in deep learning, challenges such as high intra-class variance and class imbalance limit precise vessel delineation. Most existing approaches for coronary DSA segmentation cannot address these issues. Furthermore, existing segmentation networks’ encoders do not directly generate semantic embeddings, which could enable the decoder to reconstruct segmentation masks effectively from these well-defined features. We propose a Supervised Prototypical Contrastive Loss (SPCL) that combines supervised and prototypical contrastive learning to enhance coronary DSA image segmentation. The supervised contrastive loss enforces semantic embeddings in the encoder, improving feature differentiation. The prototypical contrastive loss allows the model to focus on the foreground class while alleviating the high intra-class variance and class imbalance problems by concentrating only on the hard-to-classify background samples. We implement the proposed SPCL loss within an MSA-UNet3+: a Multi-Scale Attention-Enhanced UNet3+ architecture. The architecture integrates key components: a Multi-Scale Attention Encoder (M-encoder) and a Multi-Scale Dilated Bottleneck (MSD-Bottleneck) designed to enhance multi-scale feature extraction and a Contextual Attention Fusion Module (CAFM) designed to preserve fine-grained details while improving contextual understanding. Experiments on a private coronary DSA dataset demonstrate that MSA-UNet3+ outperforms state-of-the-art methods, achieving the highest Dice coefficient and F1-score and significantly reducing ASD and ACD. The developed framework provides clinicians with precise vessel segmentation, enabling accurate identification of coronary stenosis and supporting informed diagnostic and therapeutic decisions. The code will be released at \url{https://github.com/rayanmerghani/MSA-UNet3plus}. 
\end{abstract}




\maketitle

\section{Introduction}
Coronary artery disease (CAD) is one of the leading causes of global mortality \cite{kaba2023application, yee2024optimized}. Accurate diagnosis is critical for effective treatment; however, traditional methods that rely on visual interpretation of coronary angiograms are not only time consuming, but also suffer from inter-observer variability and human error. Digital subtraction angiography (DSA) remains the gold standard imaging modality for coronary artery disease (CAD), offering high spatial and temporal resolution \cite{kaba2023application, zhang2024cidn}.  However, extracting meaningful data from DSA images faces significant challenges, including anatomical interference (ribs, spine, diaphragm), overlapping vasculature, heterogeneous contrast distribution, and motion artifacts \cite{minaee2021image}. 

These factors affect the visibility of the coronary vessel, highlighting the need for advanced segmentation techniques to improve the diagnostic outcome. Image segmentation, defined as the process of partitioning images into distinct semantic regions \cite{minaee2021image,malathy2024deep}, represents a fundamental step in coronary artery analysis. This technique enables quantitative assessment of vessel morphology, stenosis detection, and plaque characterization \cite{raval2024medical,yang2019deep}. Although traditional methods often struggle with the complexities of DSA images, deep learning has demonstrated remarkable success in image analysis \cite{minaee2021image,chen2020deep}. Its ability to learn intricate features from large datasets makes it particularly effective for coronary artery segmentation \cite{raval2024medical,yang2019deep}.

Several studies have investigated advanced deep learning architectures and training strategies for coronary DSA image segmentation. Zhang et al. \cite{zhang2024cidn} introduced CIDN, a model for X-ray angiography segmentation, incorporating a Bio-inspired Attention Block (BAB) and a Multi-scale Interactive Block (MIB). Additionally, it combines Binary Cross-Entropy (BCE) and Adaptive Cross-Entropy (ACE) loss functions. Deng et al. \cite{deng2024dfa} proposed DFA-Net, a dual-branch network for coronary vessel segmentation in X-ray DSA images using a Contrast Improvement Enhancement Transformer (CIET) and ResUnet++ architecture, which mitigates class imbalance through joint risk cross-entropy and Dice loss. Despite outperforming existing methods, DFA-Net underutilizes temporal and spatial information. Shen et al. \cite{shen2023dbcu} developed DBCU-Net, integrating U-Net, DenseNet, and bi-directional ConvLSTM (BConvLSTM) to improve feature extraction and contextual understanding. DBCU-Net is limited by class imbalance and high computational costs. Cui et al. \cite{cui2023spatial} introduced SMAU-Net, which employs a Multi-scale Spatial Attention module, a Feature Aggregation module, and a Detail Supervision module to address complex vascular structures. Although it surpasses U-Net, SMAU-Net struggles with class imbalance and fine vessel segmentation. Zhang et al. \cite{zhang2023centerline} proposed a Centerline-Supervision Multi-Task Learning Network, improving U-Net with a Channel Attention Skip module and a Centerline Auxiliary Supervision module. Despite outperforming state-of-the-art methods, it is constrained by class imbalance and computational complexity. Zhu et al. \cite{zhu2021coronary} developed a pyramid Scene Parsing Network (PSPNet), a multi-scale CNN utilizing transfer learning to address low contrast and data scarcity. However, PSPNet faces challenges such as class imbalance and computational demands.
\begin{figure}
	\centering
	\includegraphics[width=0.8\textwidth]{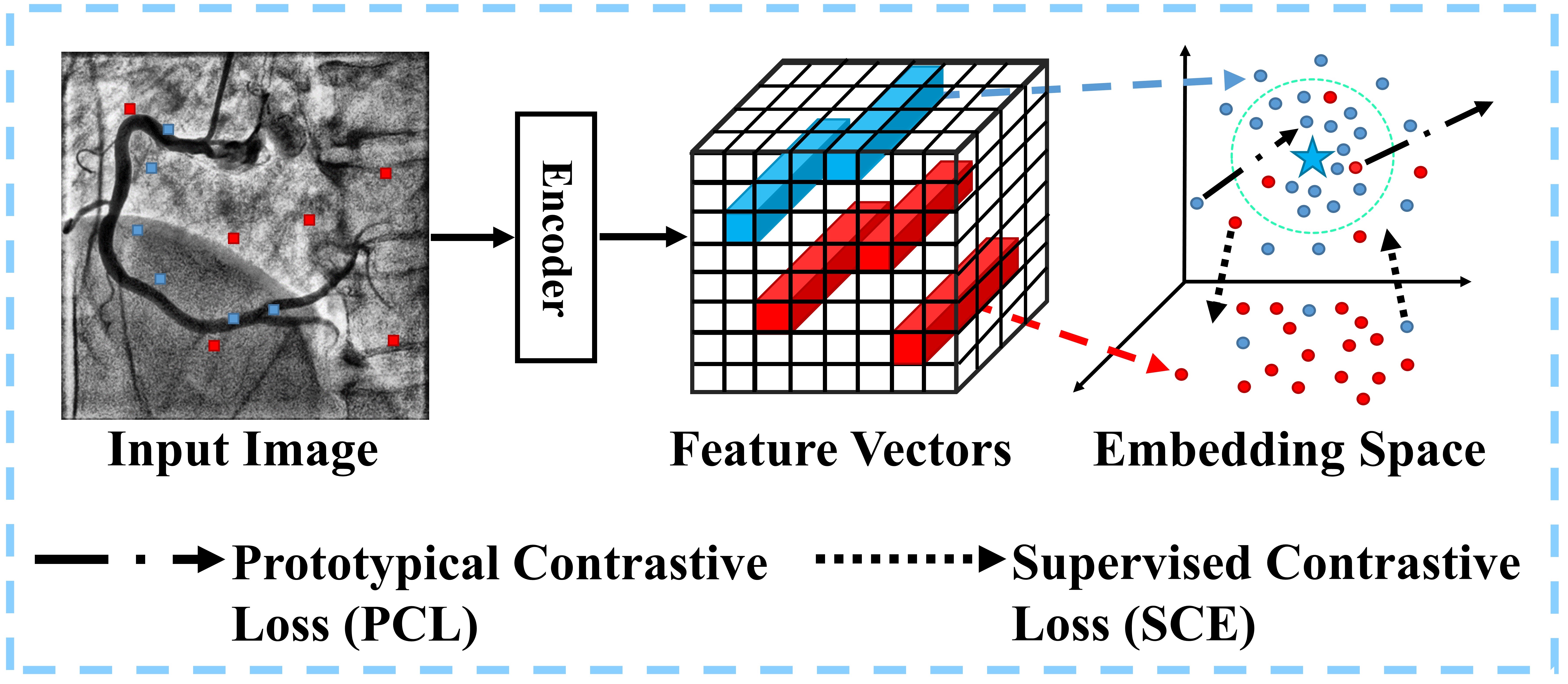}
	\caption{Illustration of the desired semantic embeddings characteristics of an encoder, which should place features from the same class close together while distancing features from different classes: SCE optimizes the embedding space by minimizing the distance between similar foreground samples (in blue) and maximizing the distance between dissimilar ones. PCL focuses on learning prototypes for foreground samples (in blue star), pulling them close to their respective prototypes while pushing hard negative instances (those close to the prototypes) further away.}
	\label{fig1}
	\vspace{-11pt} 
\end{figure}

Segmentation networks in the reviewed studies, particularly encoder-decoder architectures, face critical limitations. Encoders often prioritize performance over semantic embedding, failing to cluster similar class vectors closely in the embedding space. This challenge intensifies in coronary DSA segmentation due to the background class's structural diversity and ambiguous boundaries, causing significant intra-class variance and class imbalance. Even current approaches focusing on semantic embedding struggle with the negative class's extreme diversity. Additionally, most of these methods neglect hard negative samples, the most optimization-informative yet least abundant because of visual dissimilarity. To address these limitations, we propose a deep learning framework that integrates prototype learning \cite{nogueira2024prototypical} and supervised contrastive loss \cite{lee2021supervised}. Our hybrid approach (Fig.~\ref{fig1}) improves encoder discriminability by: (1) clustering positive features while distancing them from negatives via supervised contrastive loss, and (2) refining background embeddings through prototypical contrastive loss that isolates hard negatives by prototype distance. This dual strategy explicitly targets class imbalance and intra-class variance by focusing on challenging background samples.
\begin{figure}
	\centering
	\includegraphics[width=1.0\columnwidth]{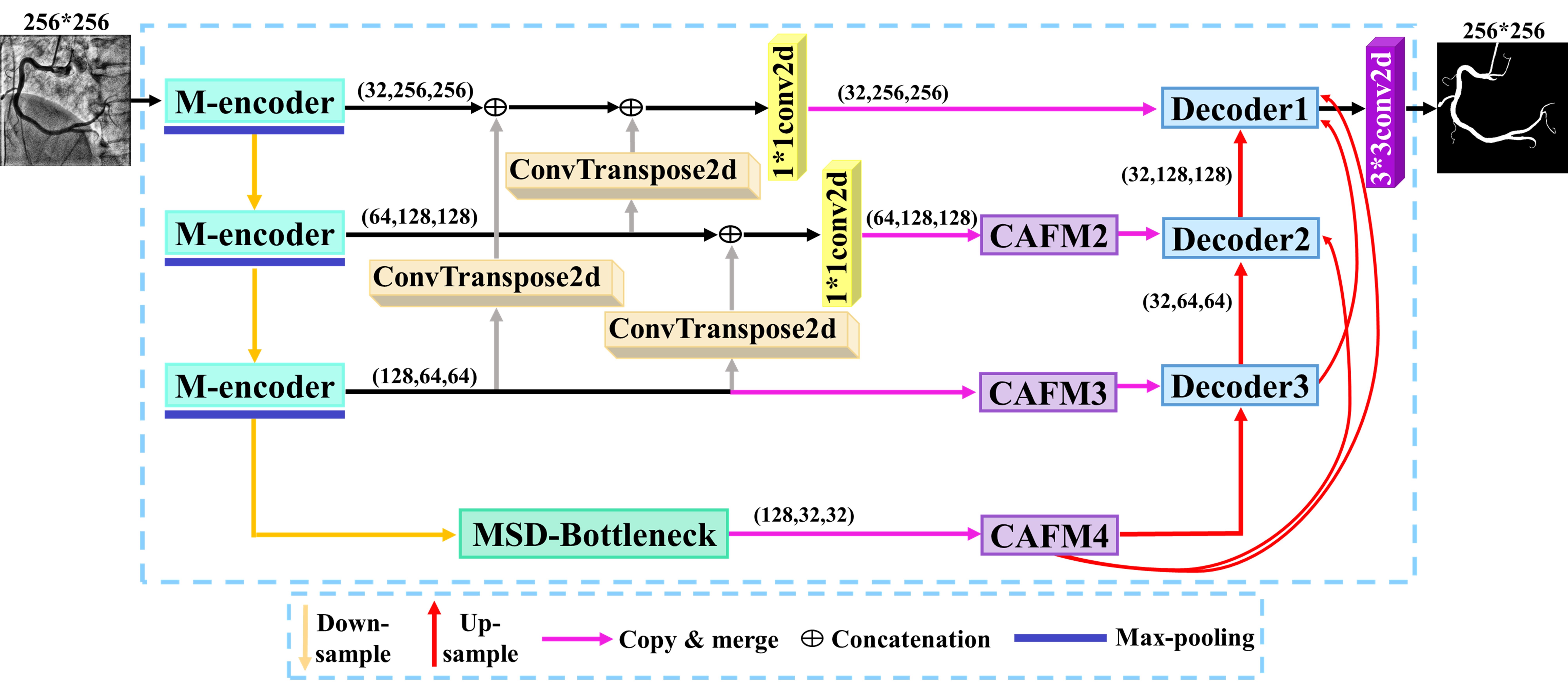}
	\caption{The architecture of the proposed MSA-UNet3+ model. The model integrates a Multi-Scale Dilated Bottleneck (MSD-Bottleneck) for multi-scale feature extraction and a Contextual Attention Fusion Module (CAFM) for enhanced contextual understanding. The M-encoder employs convolutional and transposed convolutional layers, while the decoders reconstruct the segmentation mask. This architecture enables precise segmentation of coronary arteries in DSA images by capturing both fine-grained details and broader structural information.}
	\label{fig2}
        \vspace{-10pt}
\end{figure}

Building on this, we integrate the proposed Supervised Prototypical Contrastive Loss (SPCL) within our Multi-Scale Attention-modified UNet3+ (MSA-UNet3+) framework (Fig.~\ref{fig2}). The architecture enhances UNet3+ through three key modifications: (1) a Multi-Scale Attention Encoder (M-encoder) replacing the traditional encoder for hierarchical feature extraction, (2) a Multi-Scale Dilated Bottleneck (MSD-Bottleneck) with Atrous Spatial Pyramid Pooling (ASPP) for multi-contextual feature fusion, and (3) a Contextual Attention Fusion Module (CAFM) that performs channel-wise feature recalibration. The decoder then synthesizes these refined multi-scale features to generate precise coronary DSA segmentation masks. 

The Key contributions of this work can be summarized as follows.  
\begin{itemize}
\item We propose Supervised Prototypical Contrastive Loss (SPCL), a new hybrid loss function that combines the strengths of supervised contrastive learning and prototypical contrastive learning. Unlike existing approaches, SPCL explicitly enforces ideal embedding properties in the encoder through contrastive loss, while its prototypical component targets class imbalance and  high intra-class variance by dynamically focusing on hard-to-classify background samples. This dual mechanism not only improves feature discriminability but also significantly enhances segmentation accuracy—addressing two critical limitations of current methods: (1) poor embedding behavior in encoders and (2) bias toward dominant classes. Experimental results demonstrate that SPCL outperforms state-of-the-art loss functions in coronary DSA segmentation, particularly in preserving fine vessel structures and suppressing background noise.

\item Our method uniquely emphasizes and handles hard-to-classify background samples, which are generally ignored in conventional coronary DSA segmentation approaches. Unlike existing methods that treat all background regions uniformly, our framework strategically focuses on these challenging cases through prototype-guided contrastive learning. This targeted approach achieves three key advancements: (1) substantially improved learning efficiency by concentrating computational resources on diagnostically ambiguous areas, (2) more robust segmentation performance, and (3) clinically relevant handling of class imbalance that better reflects real-world coronary imaging conditions. The resulting framework shows superior generalization capability compared to current clinical implementations.

\item We present MSA-UNet3+, a novel Multi-Scale Attention-enhanced U-Net3+ architecture that offers new standards for coronary DSA image segmentation through three key innovations: (1) a hierarchical Multi-Scale Attention Encoder that adaptively focuses on vessel structures across different resolutions, (2) a Multi-Scale Dilated Bottleneck with pyramid dilation rates (1,2,4,8) for comprehensive feature extraction, and (3) Contextual Attention Fusion Modules (CAFM) that dynamically recalibrate spatial and channel-wise features. Specifically optimized for coronary DSA challenges. The architecture's unique design enables unprecedented precision in delineating complex coronary anatomies, particularly in low-contrast regions where conventional methods typically fail.

\end{itemize}

This paper is organized as follows. \hyperref[sec:related]{Section 2} reviews recent related works. \hyperref[sec:method]{Section 3} details our proposed method. \hyperref[sec:results]{Section 4} shows the experimental results and the analysis. \hyperref[sec:discussion]{Section 5} provides a discussion of our results. \hyperref[sec:conclusion]{Section 6} concludes this work.

\section{Related work}
\label{sec:related}
Medical image segmentation is critical for computer-aided diagnosis and treatment planning \cite{xie2024weakly, xu2024advances}. While U-Net and its variants \cite{ehab2023performance, colbert2024repurposing} excel at capturing spatial contextual information. However, standard U-Net struggles with fine details. Variants like Res-UNet and Attention Res-UNet address this issue. Recent advances include (1) Convolutional Neural Networks (CNNs)-based architectures for robust feature extraction \cite{xu2024advances, abdulwahhab2024review, alom2018recurrent}, (2) Fully Convolutional Networks (FCNs) for flexible input sizes \cite{malathy2024deep}, (3) Recurrent Neural Networks (RNNs) / ConvLSTMs for temporal modeling \cite{zuo2021r2au,azad2019bi}, and (4) Generative Adversarial Networks (GANs) for data augmentation \cite{illimoottil2023recent}. Transformers (e.g., TransUNet \cite{yao2024cnn}) and foundation models such as SAM \cite{colbert2024repurposing} now enable attention-based long-range dependencies. However, class imbalance remains a fundamental challenge, particularly in medical imaging where background dominance biases model performance \cite{asif2024conceptual}.

\subsection{Traditional approaches to class imbalance}
Traditional methods for handling class imbalance in machine learning, such as data resampling techniques, can be applied to medical image segmentation. Oversampling, such as creating synthetic samples from the minority class, aims to balance the class distribution \cite{farooq2023synthetic}. However, naive oversampling methods can lead to overfitting, especially in medical images, where generating truly representative synthetic data is challenging. Undersampling, on the other hand, involves removing samples from the majority class \cite{walsh2022comparison}. Although this can reduce computational costs and mitigate overfitting, it leads to information loss and potentially discarding valuable data. The effectiveness of these techniques is highly dependent on the specific dataset and the nature of the class imbalance \cite{walsh2022comparison}. However, even with oversampling, careful consideration must be given to the methods used to generate synthetic data to ensure that they accurately represent the underlying data distribution and prevent overfitting \cite{farooq2023synthetic}.

\subsection{Loss function modifications}
Addressing class imbalance directly with the loss function is another common strategy. While standard cross entr-
opy is sensitive to class imbalance and thus favors majority classes \cite{yeung2022unified,hosseini2024dilated}, three key adaptations exist: (1) Weighted cross-entropy rebalances class contributions through manual weighting, preventing the model from being dominated by the majority class \cite{hosseini2024dilated,li2020analyzing}, However, determining appropriate weights can be challenging and requires potentially hyperparameter tuning \cite{al2024efficient}; (2) Focal loss \cite{yeung2022unified,yeung2021mixed} reduces the contribution of easily classified samples (often from the majority class) and focuses learning on more challenging samples from the minority classes; and (3) Tversky loss \cite{abraham2019novel}, a generalization of the Dice loss, offers another avenue to address class imbalance by adjusting the weights assigned to false positives and false negatives, allowing for a more balanced trade-off between precision and recall. Selection between losses depends on the dataset characteristics and desired precision-recall tradeoffs \cite{yeung2022unified}.

\subsection{Advanced techniques}
Advanced techniques beyond traditional and loss function-based approaches address the class imbalance in medical image segmentation. Semi-supervised learning, using labeled and unlabeled data, shows promise \cite{jiao2024learning,lee2024clue}. Methods like contrastive learning and consistency regularization \cite{xing2023multi,wang2023dual} learn robust representations from limited labeled data, enhancing minority class performance. These approaches involve data augmentation and enforcing consistency between predictions on augmented versions of the same image. For example, the multitask contrastive learning framework addresses class imbalance through global and local contrastive learning with multi-scale uncertainty estimation \cite{xing2023multi}. Additionally, the dual-path framework tackles intra-class imbalance by separating objects into subclasses and using weighted maps to guide segmentation \cite{lin2023dual}.

Although most contrastive learning studies focus on unsupervised settings, Khosla et al. \cite{khosla2020supervised} extended it to supervised learning by treating all samples from the same class as positive. Contrastive learning has also been applied to semantic segmentation. Chaitanya et al. \cite{chaitanya2020contrastive} introduced techniques to exploit structural similarities in volumetric medical images within the self-supervised framework, categorizing slices of similar volumes as positive pairs and dissimilar ones as negative pairs. Similarly, Zhao et al. \cite{zhao2021contrastive} proposed a fine-tuning approach using contrastive loss for semantic segmentation. However, these studies primarily aim to improve performance with limited data, rather than training segmentation encoders to inherently learn semantic embeddings. Despite significant progress, challenges remain in addressing class imbalance in medical image segmentation. Generating high-quality synthetic data is difficult, and the optimal choice of loss functions or augmentation strategies is often dataset-specific \cite{wu2024ae}. 

Overcoming these challenges is critical for accurate coronary DSA image segmentation. This study integrates supervised and prototypical contrastive learning strategies, proposing a novel hybrid loss combining supervised contrastive loss with prototypical contrastive loss. This approach enhances the encoder's ability to generate discriminative features, enabling better differentiation between diverse classes. By focusing on challenging negative samples, hybrid loss effectively addresses class imbalance and high intra-class variance.

\section{Method}
\label{sec:method}
\subsection{Overview}
In this study, we first propose the Supervised Prototypical Contrastive Loss (SPCL) to resolve coronary DSA segmentation’s core challenges: high intra-class variance from ambiguous backgrounds (ribs, spine, diaphragm, and lung) and class imbalance. SPCL uniquely combines supervised contrastive learning’s discriminative power with prototype-based hard mining for class-aware embeddings. Using SPCL, we then design MSA-UNet3+, a multi-scale attention network that achieves clinically significant vessel-background separation.

\begin{figure}
	\centering
	\includegraphics[width=0.9\columnwidth]{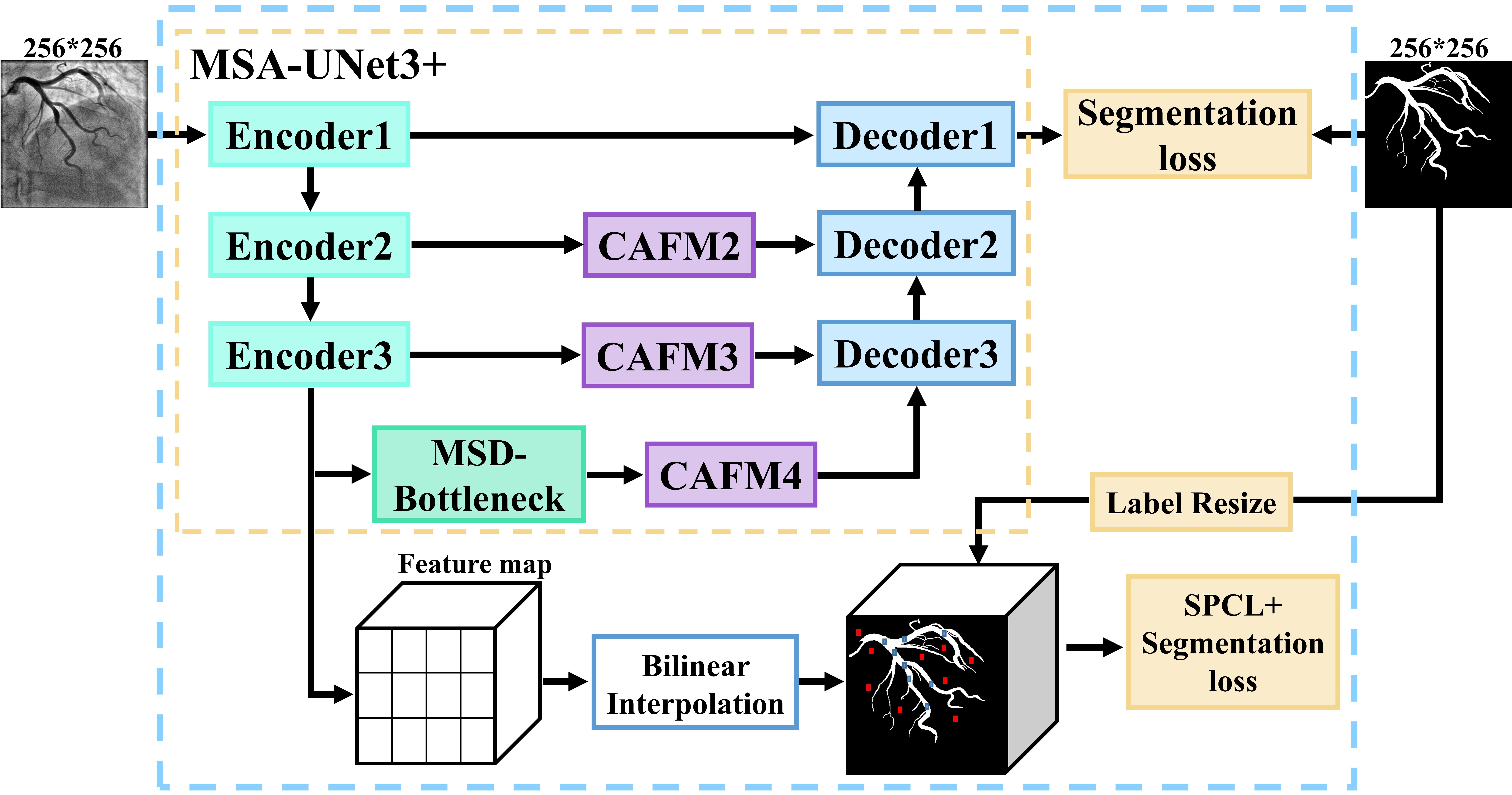}
	\caption{Architecture of the proposed MSA-UNet3+ model, highlighting the application of Supervised Prototypical Contrastive Loss (SPCL). The SPCL is applied exclusively to the final encoder output (after bilinear interpolation to 128×128 resolution), where feature representations are most semantically discriminative. The encoder extracts multi-scale features. Feature embeddings are jointly optimized via the Supervised Prototypical Contrastive Loss (SPCL) for discriminative representation learning and segmentation loss (Dice + BCE (Binary Cross Entropy)) for precise boundary delineation. CAFM denotes Contextual Attention Fusion Module, and MSD represents the Multi-Scale Dilated Bottleneck.}
	\label{fig11}
	\vspace{-11pt} 
\end{figure}

\subsection{Supervised prototypical contrastive learning}
Fig.~\ref{fig1} illustrates our proposed method pipeline, which integrates prototypical \cite{nogueira2024prototypical} and supervised contrastive learning \cite{lee2021supervised}. The framework consists of two key components: 1) Supervised Contrastive Embedding (SCE) learns discriminative feature representations by minimizing distances between feature vectors of semantically similar regions (foreground-foreground or background-background pairs) while maximizing distances between dissimilar regions (foreground-background pairs). This ensures that semantically related image regions cluster in the embedding space. 2) Prototypical Contrastive Loss (PCL) enhances feature separation by attracting foreground samples to class-specific prototypes and repelling background samples from these prototypes. This reduces the training burden by focusing on distinguishing foreground from background, which exhibits high intra-class variability. In addition, the model prioritizes challenging background samples based on their distance from foreground prototypes, enhancing the discriminative capability. To this end, we use the resized ground-truth masks for supervision as depicted in Fig.~\ref{fig11}.

\vspace{5pt}
We combine these two components into a unified Supervised Prototypical Contrastive Loss (SPCL) that is applied to feature embeddings extracted from the final encoder layer of MSA-UNet3+. This strategic choice is based on the observation that the final encoder output contains the most semantically abstract and consolidated feature representations, having processed information through all preceding convolutional and attention layers, and benefits from the largest receptive field which enables global understanding essential for distinguishing coronary vessels from complex anatomical backgrounds. These high-level features provide optimal discriminative power for both the supervised and prototypical contrastive learning components of SPCL, offering optimal class-specific information for effective contrastive learning. The embeddings are resized to $128 \times 128$ resolution via bilinear interpolation before SPCL computation as illustrated in Fig.~\ref{fig11}, ensuring consistent embedding dimensions. Applying SPCL at multiple encoder levels would introduce significant computational overhead with diminishing returns, as discriminative power naturally concentrates toward the network's deepest layers, making the final encoder output the most efficient and effective point of application for our contrastive learning framework.

In implementing the PCL component of SPCL, hard negative samples are identified using a distance-to-prototype thresholding mechanism. Specifically, for each background sample, we compute the cosine distance $D(z_i, p_k)$ between its feature embedding and the learned foreground class prototype. A background sample is classified as `hard' if its distance to the prototype is less than the margin parameter $m$, which defines a hypersphere around the prototype in the embedding space. This thresholding mechanism uses an absolute distance threshold rather than top-k selection and is implemented through the term $\max(0, m - D(z_i, p_k))$ in Eq.~(\ref{eq8}) (section 3.3.4). Background samples inside this region ($\text{distance} < m$) are considered challenging because they are semantically close to the foreground class and thus require explicit separation. When $D(z_i, p_k) < m$, this term activates, applying a repulsive force to push the sample away from the prototype. Samples outside this region ($\text{distance} \geq m$) are treated as `easy negatives' that are already well-separated and receive minimal gradient updates, as the term becomes zero when $D(z_i, p_k) \geq m$. This threshold-based approach directly addresses the high intra-class variance in coronary DSA backgrounds by concentrating learning efforts only on those background samples that are genuinely difficult to distinguish from coronary vessels, and it is computationally efficient, requiring only comparison operations rather than distance sorting.

The combined Supervised Prototypical Contrastive Loss (SPCL) leverages both global (SCE) and local (PCL) feature relationships. By adaptively weighting hard negative samples based on their distance to foreground prototypes, SPCL addresses class imbalance and intra-class variability. This dual strategy yields robust feature representations that improve segmentation performance, particularly for challenging cases with heterogeneous background structures.

\subsection{Architecture overview and loss function}

U-Net remains foundational for medical image segmentation. Advances like U-Net++ and U-Net3+ improved this baseline. U-Net3+ excels via an encoder-decoder structure with interconnections and intraconnections, preserving spatial information while capturing multi-scale features. It achieves superior learning efficiency with reduced computational overhead. Building on these strengths, we employ U-Net3+ as the backbone for the proposed MSA-UNet3+.

MSA-UNet3+ extends U-Net3+ for medical image segmentation, optimized for coronary DSA. As shown in Fig.~\ref{fig2}, it uses an encoder-decoder structure with three innovations: First, the multi scale attention encoder (M-encoder) instead of normal encoder performs hierarchical downsampling to extract multi-scale features. Second, the multi-scale dilated bottleneck (MSD-Bottleneck) layer integrates the Atrous Spatial Pyramid Pooling (ASPP) module for multi-context feature aggregation. Third, Contextual Attention Fusion Modules (CAFM) combine encoder outputs through multi-feature fusion, combining fine details with semantic context. The architecture generates two outputs: (1) a segmentation mask through the final convolutional layer and (2) a compact feature representation via an embedding layer for auxiliary analysis as in Fig Fig.~\ref{fig11}. Together, these design elements collectively enable MSA-UNet3+ to achieve state-of-the-art performance in challenging coronary DSA segmentation tasks, particularly for complex anatomical structures affected by noise and low contrast.

\begin{figure}
	\centering
	\includegraphics[width=0.8\columnwidth]{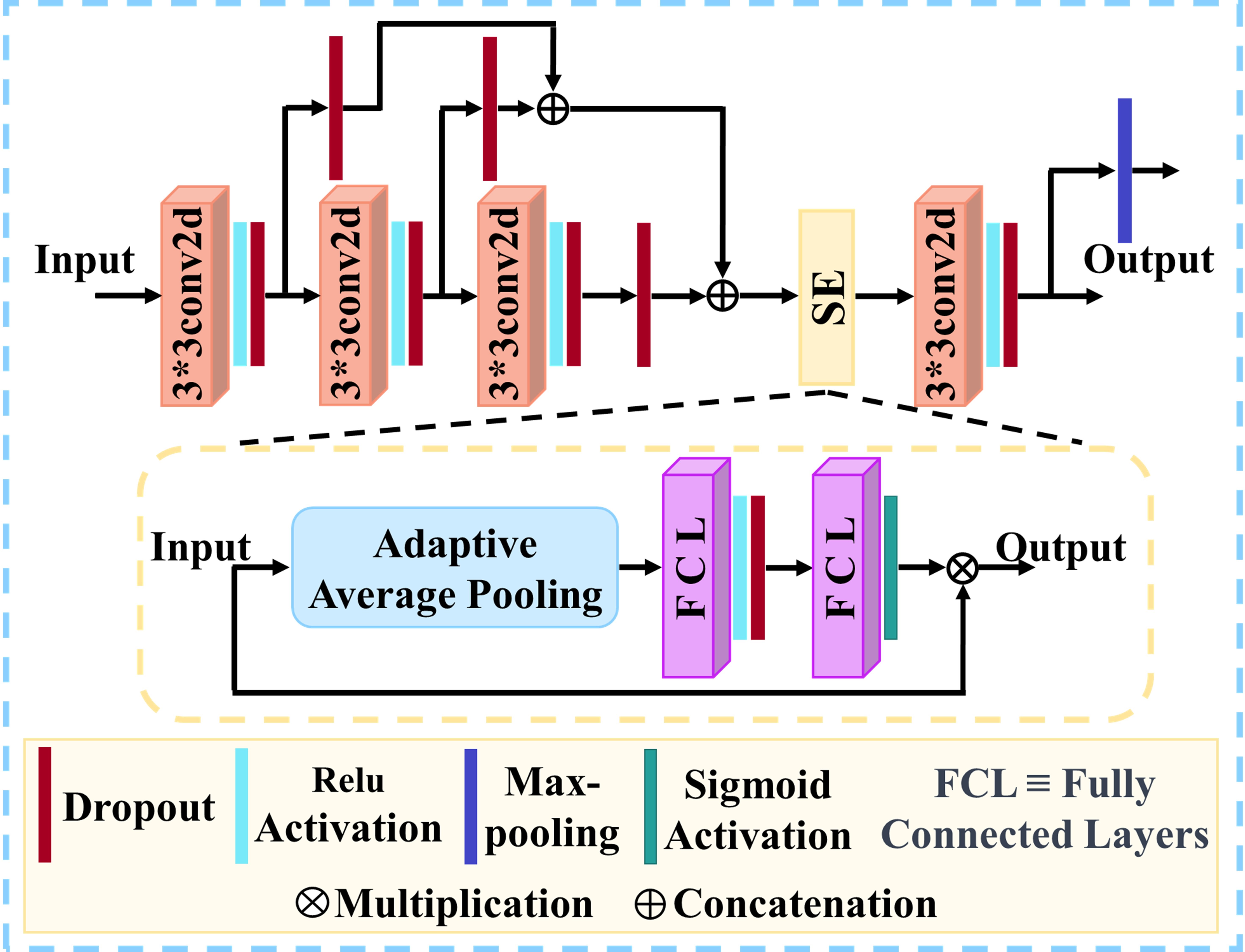}
	\caption{The architecture of the M-Encoder (Multi-Scale Attention module) highlights the Squeeze-and-Excitation (SE) block implementation.}
	\label{fig3}
        \vspace{-10pt}
\end{figure}

\subsubsection{Multi-scale attention encoder (M-Encoder)}
Traditional encoders face limitations in feature extraction, spatial preservation, and contextual understanding. We propose an M-Encoder (Fig.~\ref{fig3}) inspired by \cite{zhou2023dual}, using three progressive convolutional layers to refine and concatenate features for deeper and more comprehensive feature extraction. A key innovation is integrating a Squeeze-and-Excitation (SE) block, which serves as a channel-wise attention mechanism that dynamically recalibrates channels, enhancing critical features while suppressing less informative ones. By concatenating multi-scale features before the SE block, the M-Encoder achieves richer representations at each layer, followed by dimensionality-reduction convolution for computational efficiency. The complete M-Encoder operation is formalized in Eq.~(\ref{eq1}):
\begin{equation}\label{eq1} 
\begin{aligned}
&\text{out}_1 = \text{conv}_{3 \times 3}(x), \\
&\text{out}_2 = \text{conv}_{3 \times 3}(\text{out}_1), \\
&\text{out}_3 = \text{conv}_{3 \times 3}(\text{out}_2), \\
&\text{Output} = \text{conv}_{3 \times 3}\big(\text{SE}\big(\text{Cat}\big(\text{out}_1, \text{out}_2, \text{out}_3\big)\big)\big)
\end{aligned}
\end{equation}
where, $x$ represents the input feature maps to the M-Encoder, $Output$ represents the output feature maps, $Cat$ denotes the concatenation operation in the channel dimension, and $SE$ represents the Squeeze-and-Excitation blocks.
\begin{figure}
	\centering
	\includegraphics[width=0.8\columnwidth]{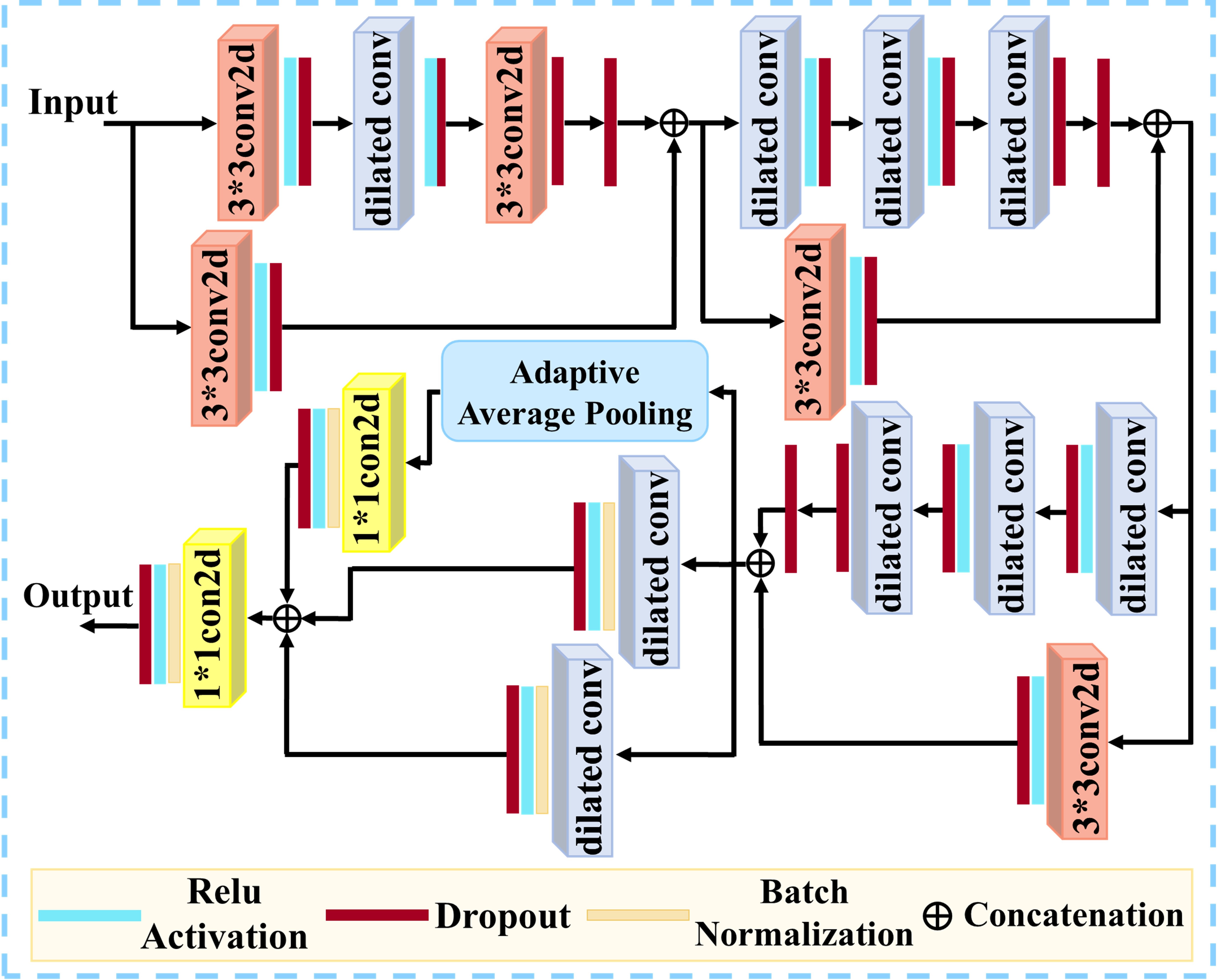}
	\caption{Detailed architecture of the Multi-Scale Dilated Bottleneck (MSD-Bottleneck) module in the proposed MSA-UNet3+ model.}
	\label{fig4}
        \vspace{-12pt}
\end{figure}
\subsubsection{Multi-scale dilated bottleneck (MSD-Bottleneck)}
Continuous convolution and pooling operations frequently degrade fine-grained details crucial for segmentation, while conventional up-sampling fails to recover them. We address this with our MSD-Bottleneck (Fig.~\ref{fig4}), building upon \cite{zhou2023dual}, combining a bottleneck structure with an Atrous Spatial Pyramid Pooling (ASPP) to preserve fine details while capturing multi-scale context. It uses three dilation patterns ([1,2,1], [2,4,2], [4,8,4]) to extract multi-scale features efficiently. Residual connections prevent vanishing gradients, while ASPP's parallel dilated convolutions (rates=[4,8]) and global average pooling enhance context. Features are concatenated and projected to the original channel dimension, combining local precision with global context. For the input $x$, processing follows Eq.~(\ref{eq2}): first through dilated bottleneck blocks with residuals, then ASPP.
\vspace{-3pt}
\begin{equation}\label{eq2}
\begin{aligned}
&\text{out}_1 = \text{conv}_{3 \times 3}(x; \text{dilation}=[1, 2, 1]), \\
&\text{out}_2 = \text{conv}_{3 \times 3}(\text{out}_1; \text{dilation}=[2, 4, 2]), \\
&\text{out}_3 = \text{conv}_{3 \times 3}(\text{out}_2; \text{dilation}=[4, 8, 4]), \\
&\text{out}_4 = \text{Cat}\Big((\text{out}_3; \text{dilation}=[4, 8]), \text{conv}_{1 \times 1}(\text{AAP}(\text{out}_3))\Big), \\
&\text{ASPP} =\text{conv}_{1 \times 1}(\text{out}_4)
\end{aligned}
\end{equation}
\subsubsection{Contextual attention fusion module  (CAFM)}
\begin{figure}
	\centering
	\includegraphics[width=0.8\columnwidth]{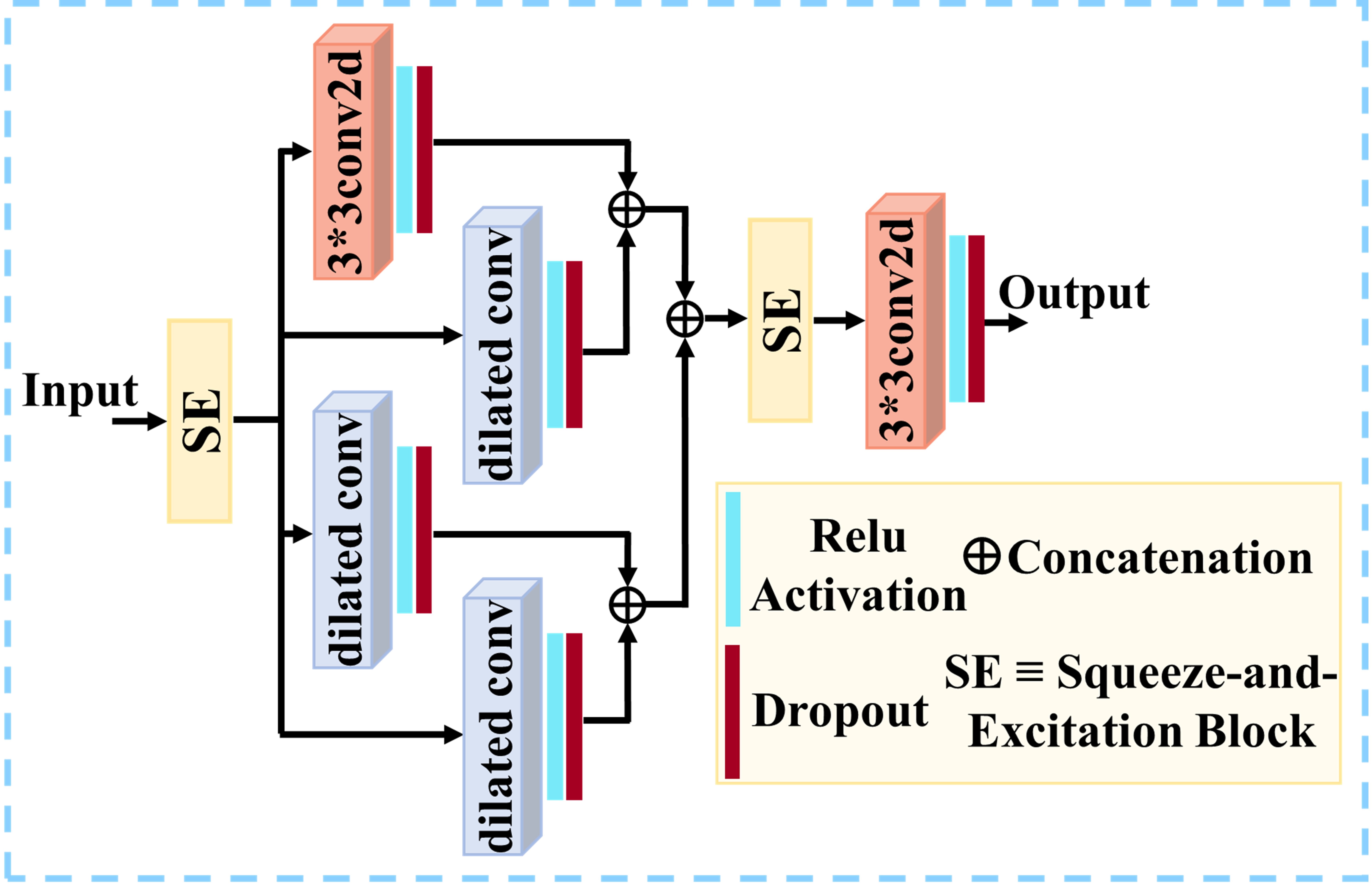}
	\caption{The architecture of the Contextual Attention Fusion Module (CAFM) in the MSA-UNet3+ model, combines Squeeze-and-Excitation (SE) blocks with multi-scale processing.}
	\label{fig5}
        \vspace{-10pt}
\end{figure}

Encoders in segmentation networks lose spatial context in deeper layers, particularly in complex medical imaging such as coronary digital subtraction angiography (DSA). Our Contextual Attention Fusion Module (CAFM) (Fig.~\ref{fig5}) addresses this by multi-scale dilated convolutions (rates=[1,2,4,8]) to expand receptive fields while preserving resolution, enabling robust feature extraction across varying vessel sizes; a key requirement in coronary DSA. Furthermore, CAFM incorporates Squeeze-and-Excitation (SE) blocks to dynamically recalibrate channel-wise feature importance, prioritizing diagnostic features. CAFM provides the decoder with refined, context-rich inputs, bridging encoder-decoder semantic gaps through enhanced multi-scale representations, this improves segmentation of fine and coarse structures. Let $x$ be the input feature map to the CAFM module. The operations performed by the module can be described as follows: 
\begin{equation}\label{eq4}
\begin{aligned}
&\text{out}_{\text{final}} = \text{conv}_{3 \times 3}\Big(SE\big(Cat(\text{x}; \text{dilation}=[1,2,4, 8])\big) \Big)
\end{aligned}
\end{equation}
where, $x$ represents the input feature maps to the CAFM module, $SE$ represents the Squeeze-and-Excitation blocks, the notation dilation=[1, 2, 4, 8] specifies four convolutional layers with dilation rates of 1, 2, 4, and 8, respectively.
\subsubsection{Loss function}
In this study, we optimize our deep learning model for coronary DSA image segmentation using a composite loss function addressing class imbalance and intra-class variability. While binary cross-entropy (BCE) and Dice loss are commonly employed, they do not inherently enforce semantically meaningful or discriminative feature learning in the encoder. To address this limitation, we combine them with the proposed novel supervised prototypical contrastive loss (SPCL), which encourages the encoder to generate semantically rich and discriminative representations, thereby enhancing the overall performance of the segmentation network.
\begin{enumerate}
\itemsep=0pt
\item Binary cross-entropy loss (BCE) \\
It measures the dissimilarity between the predicted probability map and the ground truth labels. BCE loss ensures the model learns to classify each pixel correctly by penalizing deviations from the ground truth. However, it may struggle with class imbalance, as it treats all pixels equally, regardless of class distribution. For a binary segmentation task, the BCE loss is defined as in Eq. ~(\ref{eq5}):
\begin{equation} \label{eq5}
\mathcal{L}_{\mathrm{BCE}}=-\frac{1}{N} \sum_{i=1}^N\left[y_i \log \left(\hat{y}_i\right)+\left(1-y_i\right) \log \left(1-\hat{y}_i\right)\right]
\end{equation}
where \( y_{i} \) is the ground truth label (0 or 1) for pixel \( i \), \( \hat{y}_{i} \) is the predicted probability of pixel \( i \) belonging to the foreground class, and \( N \) is the total number of pixels. 

\item Dice loss \\
Dice loss quantifies the similarity between the predicted and ground truth segmentation masks. By maximizing this overlap, it directs the model's focus to regions of interest, enhancing segmentation performance. Dice loss complements BCE by prioritizing accurate foreground class predictions, essential for precise segmentation in coronary DSA images. The Dice loss is defined in Eq. ~(\ref{eq6}):
\begin{equation}\label{eq6}
\mathcal{L}_{\text {Dice }}=1-\frac{2 \sum_{i=1}^N y_i \hat{y}_i}{\sum_{i=1}^N y_i+\sum_{i=1}^N \hat{y}_i}
\end{equation}
where \( y_{i} \) and \( \hat{y}_{i} \) are the ground truth and predicted probability for pixel \( i \), respectively. 
\item Supervised prototypical contrastive loss \\
To address class imbalance and high intra-class variance, we introduce the Supervised Prototypical Contrastive Loss (SPCL). This loss combines supervised and prototypical contrastive losses, enhancing the encoder's ability to differentiate between diverse classes. 
(1) Supervised Contrastive Loss: This loss encourages the encoder to map feature vectors of similar image samples (foreground or background regions) closer in the embedding space while pushing dissimilar ones apart. The loss function is defined as in Eq.~(\ref{eq7}):
\vspace{-3pt}
\begin{equation} \label{eq7}
L_{\mathrm{SCE}}=\sum_{i=1}^N \sum_{j \in P(i)} \log \frac{\exp \left(\operatorname{sim}\left(f_i, f_j\right) / \tau\right)}{\sum_{k \in A(i)} \exp \left(\operatorname{sim}\left(f_i, f_k\right) / \tau\right)}
\end{equation}
where \( f_{i} \) and \( f_{j} \) are feature vectors, \( P(i) \) is the set of positive pairs for \( i \), \( A(i) \) is the set of all pairs for \( i \), \( \text{sim} \) is a cosine similarity function, and \( \tau \) is a temperature parameter set to 1 without hyperparameter tuning.

(2) Prototypical Contrastive Loss: This loss learns foreground class prototypes, pulling foreground samples toward them while pushing background samples away. The loss function is defined as in Eq.~(\ref{eq8}):
\vspace{-3pt}
\begin{equation} \label{eq8}
\begin{split}
\mathcal{L}_{\mathrm{PCL}} &= \frac{1}{N \cdot n_{\mathrm{p}}} \sum_{i=1}^N \sum_{k=1}^{n_{\mathrm{p}}} \biggl[w_1 \cdot y_i \cdot D\left(z_i, p_k\right)^2 \\
&\quad + w_0 \cdot \left(1 - y_i\right) \cdot \max \left(0, m - D\left(z_i, p_k\right)\right)^2 \biggr]
\end{split}
\end{equation}
where \( N \): Total number of valid embeddings (pixels), \( n_{\text{p}} \): Number of prototypes set to 2 without hyperparameter tuning, \(D\,(z_{i}, p_{k}) \): Cosine distance between embedding \( z_{i} \) and prototype \( p_{k} \), \( w_{1} \): Weight for positive samples (close to their prototype), \( w_{0} \): Weight for negative samples (far from their prototype), and \( m \): Margin parameter controlling class separation.

\textbf{Margin determination:}

The margin parameter \( m \) in Eq.~(\ref{eq8}) is empirically determined for each architecture during preliminary experiments and falls within the range (3.5 to 8.5) based on cosine distance in the normalized embedding space. For our proposed MSA-UNet3+, we used \( m \) = 6.3, it's also remained constant throughout training for that model. The value of \( m \) is calculated based on the distance distributions between pixel embeddings and the foreground prototype. Specifically, we analyze \( D_{\text{pos}} \) (distances of positive samples to prototype) and \( D_{\text{neg}} \) (distances of negative samples to prototype), setting \( m \) to optimally separate these distributions. In practice, we optimize \( m \) separately for each model architecture through empirical analysis of distance distributions. This adaptive approach ensures that the model focuses on hard-to-classify samples while automatically adjusting to varying image complexities.

\textbf{Prototype initialization and update:}

The prototypes are initialized as learnable parameter vectors with random values sampled from a uniform distribution. We use $n_p = 2$ prototypes for the foreground class, each with dimensionality matching the encoder's output embedding space. These prototypes are treated as trainable parameters from the start of training. During training, the prototypes are updated via gradient descent through the SPCL loss using the same Adam optimizer as all other network parameters. The update pulls positive samples toward the prototypes and pushes hard negative samples (those within distance $m$) away from the prototypes. This update strategy allows the prototypes to converge to positions that minimize intra-class variance for vessels while maximizing separation from challenging background structures, and enables them to continuously adapt to the evolving feature space throughout training. Through empirical experimentation with 1, 2, 3, and 4 prototypes, we found that two prototypes provided the optimal balance between representational capacity and model complexity. This allows the model to potentially capture different semantic substructures within the foreground class---such as main vessel trunks versus fine branches---without risking overfitting.

\vspace{5pt}
(3) Total Loss: The total loss is a weighted combination of the aforementioned losses, as defined in Eq.~(\ref{eq9}):
\begin{equation} \label{eq9}
L_{\mathrm{total}}=\alpha L_{\mathrm{BCE}}+\beta L_{\mathrm{Dice}}+\gamma L_{\mathrm{SPCL}}
\end{equation}
where, \( \alpha \), \( \beta \), and \( \gamma \) are hyperparameters controlling the contribution of each loss component, all set to 1 without hyperparameter tuning. This combined loss leverages the strengths of individual losses to improve segmentation performance in coronary DSA image tasks, with L$_{\text{SPCL}}$ combining L$_{\text{SCE}}$ and L$_{\text{PCL}}$.
\end{enumerate} 

\section{Experimental results}
\label{sec:results}
\subsection{Dataset }
The DSA dataset used in this study was acquired from the authors of \cite{pu2023semi}, with original data provided by the General Hospital of Southern Theater Command. The dataset consists of 300 coronary X-ray angiographic sequences from 50 patients, each with a resolution of \(512 \times 512\) pixels, including both left and right coronary imaging sessions. For this study, we utilized only the right coronary artery images (150 images), which capture contrast agent flow through the coronary vasculature. Clinicians selected the most diagnostically relevant frames, which were then resized to 256$\times$256 pixels during pre-processing. The dataset was divided into 120 training and 30 testing images. We employed 5-fold cross-validation on the training set, with results averaged across all folds for final evaluation.
\subsection{Implementation details}
The proposed model was implemented using the PyTorch and PyTorch Lightning frameworks, enabling flexible and reproducible deep learning experimentation. All experiments were conducted on an NVIDIA RTX 4090 GPU (24GB VRAM). Each image and its corresponding label were resized to a fixed resolution of 256×256 pixels. The dataset, comprising $256 \times 256$ pixel images and corresponding labels, was normalized to the range $[-1, 1]$ using a mean and standard deviation of 0.5 \cite{10552074}. During training, we applied data augmentation techniques, including random brightness \cite{wu2023image} and contrast adjustments \cite{wu2025adaptive}, to improve model generalization. Optimization was performed using the Adam optimizer with an initial learning rate of $1 \times 10^{-4}$, decayed by a factor of 0.1 at 60\% and 80\% of the total 100 training epochs. A batch size of 5 was maintained throughout training.

\subsection{Evaluation metrics} 
We evaluated segmentation models using five metrics: recall, F1-score, Dice coefficient, average contour distance (ACD), and average surface distance (ASD).
\begin{enumerate}
\itemsep=0pt
\item Dice coefficient (Dice Similarity Coefficient, DSC) \\
It measures the overlap between the predicted segmentation mask and the ground truth, it emphasizes correct foreground class predictions. The mathematical representation of DSC is illustrated in Eq. (~\ref{eq10}):
\vspace{-2pt}
\begin{equation} \label{eq10}
\text{Dice} = \frac{2 \cdot \text{TP}}{2 \cdot \text{TP} + \text{FP} + \text{FN}}
\end{equation}
where True Positives (TP): Pixels correctly predicted as foreground; False Positives (FP): Pixels incorrectly predicted as foreground; and False Negatives (FN): Pixels incorrectly predicted as background.

\item Recall (Sensitivity) \\
It measures the proportion of true positive pixels correctly identified by the model, as defined in Eq.~ (\ref{eq11}):
\vspace{-2pt}
\begin{equation} \label{eq11}
\text{Recall} = \frac{TP}{TP + FN}
\end{equation}
Recall is critical metrics in medical imaging, as missing parts of the target structure (e.g., coronary arteries) can have serious consequences.

\item F1-score \\
It is the harmonic mean of precision and recall, provides a balanced measure of model performance, as defined in Eq. ~(\ref{eq12}):
\vspace{-2pt}
\begin{equation} \label{eq12}
\text{F1-score} = \frac{2 \times (\text{precision} \times \text{recall})}{\text{precision} + \text{recall}}
\end{equation}
Precision measures the proportion of correctly predicted foreground pixels (\(\text{Precision} = \frac{TP}{FP + TP}\)). The F1-score balances precision and recall, making it a comprehensive metric particularly in imbalanced datasets.

\item Average surface distance (ASD) \\
ASD measures the average distance between the surfaces of the predicted and ground truth masks, as defined in Eq.~ (\ref{eq13}):
\vspace{-2pt}
\begin{equation} \label{eq13}
\mathrm{ASD} = \frac{1}{2} \left( \frac{1}{|S_P|} \sum_{p \in S_p} d(p,S_G) + \frac{1}{|S_G|} \sum_{g \in S_G} d(g,S_P)\right)
\end{equation}
where, \( S_{P} \) and \( S_{G} \) are the surfaces of the predicted and ground truth masks, respectively; \( d(p,S_{G}) \) is the minimum Euclidean distance from point \( p \) on \( S_{P} \) to \( S_{G} \); and \( d(g,S_{P}) \) is the minimum Euclidean distance from point \( g \) on \( S_{G} \) to \( S_{P} \). 

ASD evaluates segmentation boundary accuracy, with lower values indicating better alignment between predicted and ground truth contours, crucial for tasks such as coronary artery segmentation.

\item Average contour distance (ACD) \\
ACD measures the average distance between the contours of the predicted and ground truth masks, specifically focusing on contour points. It is defined as in Eq.~(\ref{eq14}):
\begin{equation} \label{eq14}
\mathrm{ACD} = \frac{1}{2} \left( \frac{1}{|C_P|} \sum_{p \in C_P} d(p,C_G) + \frac{1}{|C_G|} \sum_{g \in C_G} d(g,C_P) \right)
\end{equation}
where, \( C_{P} \) and \( C_{G} \) are the contours of the predicted and ground truth masks, respectively; \( d(p,C_{G}) \) is the minimum Euclidean distance from point \( p \) on \( C_{P} \) to \( C_{G} \); and \( d(g,C_{P}) \) is the minimum Euclidean distance from point \( g \) on \( C_{G} \) to \( C_{P} \). 

ACD provides a focused evaluation of contour alignment, critical for tasks where precise shape and boundaries are essential, such as vascular structure segmentation.
\end{enumerate}  

\begin{table*}[htbp]
  \centering
  \caption{Segmentation performance comparison on coronary DSA dataset assessing SPCL effectiveness. Metrics shown for six architectures at $\gamma = 0$ and $\gamma = 1$, with 5-fold cross-validation results (means $\pm$ standard deviations).}
  \begin{tabular}{l l l l l l l}
    \toprule
    \multicolumn{1}{l}{\textbf{Architecture}} & \multicolumn{1}{c}{\textbf{$\gamma$}} & \multicolumn{1}{c}{\textbf{Recall $\uparrow$}} & \multicolumn{1}{c}{\textbf{F1 $\uparrow$}} & \multicolumn{1}{c}{\textbf{Dice $\uparrow$}} & \multicolumn{1}{c}{\textbf{ASD $\downarrow$}} & \multicolumn{1}{c}{\textbf{ACD $\downarrow$}} \\
    \midrule
    \multirow{2}[0]{*}{\textbf{U-Net \cite{ronneberger2015u}}} & \multicolumn{1}{c}{0} & 85.79±0.0080 & 87.38±0.0054 & 87.33±0.0054 & 0.80±0.0456 & 0.78±0.0407 \\[0.1cm]
          & \multicolumn{1}{c}{1}  & \textbf{86.24±0.0056}$\diamond$$\diamond$ & \textbf{87.91±0.0034}$\nabla$ & \textbf{87.87±0.0033}$\nabla$ & \textbf{0.77±0.0602}$\diamond$$\diamond$ & \textbf{0.75±0.0563}$\diamond$$\diamond$ \\[0.1cm]
    \multirow{2}[0]{*}{\textbf{AU-Net \cite{oktay2018attention}}} & \multicolumn{1}{c}{0} & 85.86±0.0084 & 87.53±0.0025 & 87.49±0.0027 & 0.82±0.06716 & 0.79±0.0595 \\[0.1cm]
          & \multicolumn{1}{c}{1}  & \textbf{86.41±0.0089}$\diamond$$\diamond$ & \textbf{87.90±0.0034}$\nabla$ & \textbf{87.85±0.0033}$\nabla$ & \textbf{0.76±0.0453}$\diamond$ & \textbf{0.74±0.0404}$\diamond$ \\[0.1cm]
    \multirow{2}[0]{*}{\textbf{UNet++  \cite{zhou2018unet++}}} & \multicolumn{1}{c}{0}  & 84.47±0.0222 & 86.86±0.0089 & 86.81±0.0090 & 0.82±0.0645 & 0.80±0.0555 \\[0.1cm]
          & \multicolumn{1}{c}{1} & \textbf{86.20±0.0105}$\diamond$ & \textbf{87.87±0.0039}$\nabla$ & \textbf{87.83±0.0039}$\nabla$ & \textbf{0.74±0.0371}$\nabla$ & \textbf{0.72±0.0351}$\nabla$ \\[0.1cm]
    \multirow{2}[0]{*}{\textbf{R2U-Net \cite{alom2018recurrent}}} & \multicolumn{1}{c}{0}  & 81.98±0.0394 & 86.28±0.0135 & 86.21±0.0136 & 1.09±0.2254 & 1.04±0.1811 \\[0.1cm]
          & \multicolumn{1}{c}{1} & \textbf{85.02±0.0108}$\diamond$ & \textbf{87.09±0.0056}$\diamond$ & \textbf{87.03±0.0056}$\diamond$ & \textbf{0.95±0.0906}$\diamond$ & \textbf{0.9093±0.0712}$\diamond$ \\[0.1cm]
    \multirow{2}[0]{*}{\textbf{NAUNet \cite{li2020attention}}} & \multicolumn{1}{c}{0}  & 85.46±0.0081 & 87.20±0.0048 & 87.15±0.0047 & 0.79±0.0618 & 0.77±0.0557 \\[0.1cm]
          & \multicolumn{1}{c}{1} & \textbf{86.55±0.0063}$\nabla$ & \textbf{87.83±0.0039}$\nabla$ & \textbf{87.79±0.0039}$\nabla$ & \textbf{0.75±0.0544}$\diamond$ & \textbf{0.73±0.0504}$\diamond$ \\[0.1cm]
    \multirow{2}[1]{*}{\textbf{R2AU-Net \cite{zuo2021r2au}}} & \multicolumn{1}{c}{0}  & 83.44±0.0116 & 85.79±0.0056 & 85.68±0.0061 & 1.27±0.16826 & 1.20±0.13306 \\[0.1cm]
          & \multicolumn{1}{c}{1} & \textbf{85.72±0.0230}$\nabla$ & \textbf{86.86±0.0057}$\nabla$$\nabla$ & \textbf{86.80±0.0057}$\nabla$$\nabla$ & \textbf{0.97±0.1568}$\nabla$$\nabla$ & \textbf{0.93±0.1248}$\nabla$$\nabla$ \\[0.1cm]
    \bottomrule
  \end{tabular}%
  \label{table1}%
  \vspace{0.1cm} 
  \begin{minipage}{\textwidth}
    \small 
    \textbf{Notes:}  
    $\nabla$ \(p-value < 0.05\), $\nabla$$\nabla$ \(p-value < 0.01\), $\diamond$ \(p-value < 0.1\), $\diamond$$\diamond$ \(p-value < 0.2\) compared with the $\gamma=0$: Baseline, $\gamma=1$: With SPCL, AU-Net \text{---} Attention U-Net \text{and} NAUNet \text{---} Attention UNet++.
  \end{minipage}
  \vspace{-8pt}
\end{table*}

\begin{figure}
	\centering
	\includegraphics[width=0.8\columnwidth]{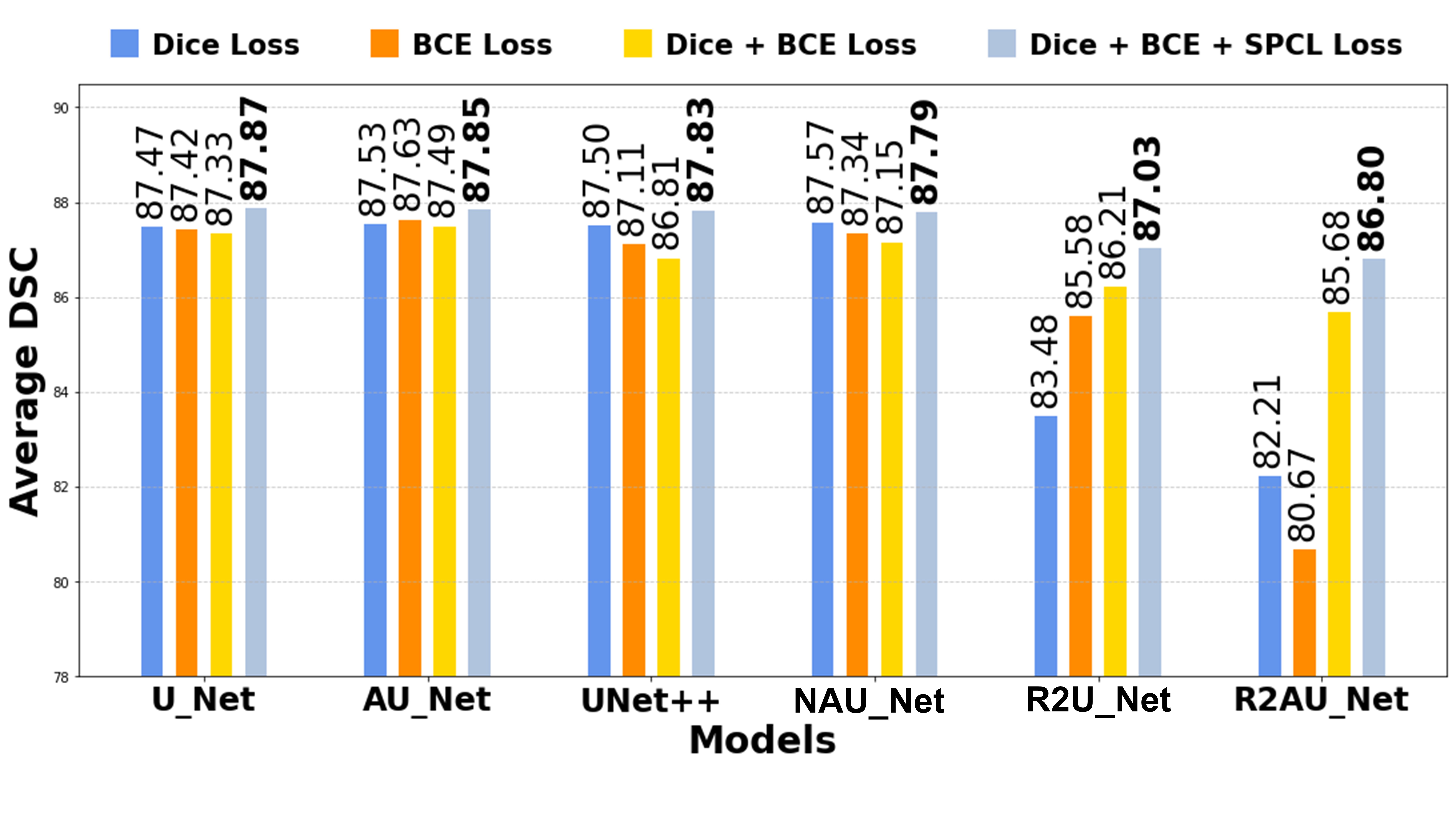}
	\caption{Comparison of average Dice Similarity Coefficient (DSC) values across six segmentation architectures using four loss functions. The proposed Dice+BCE+SPCL combination consistently achieves the highest performance for all models, demonstrating the additive benefit of Supervised Prototypical Contrastive Loss (SPCL). Numerical labels indicate absolute DSC (\%) values and improvement margins.}
	\label{fig9}
        \vspace{-8pt}
\end{figure}

\begin{figure}
	\centering
	\includegraphics[width=0.9\columnwidth]{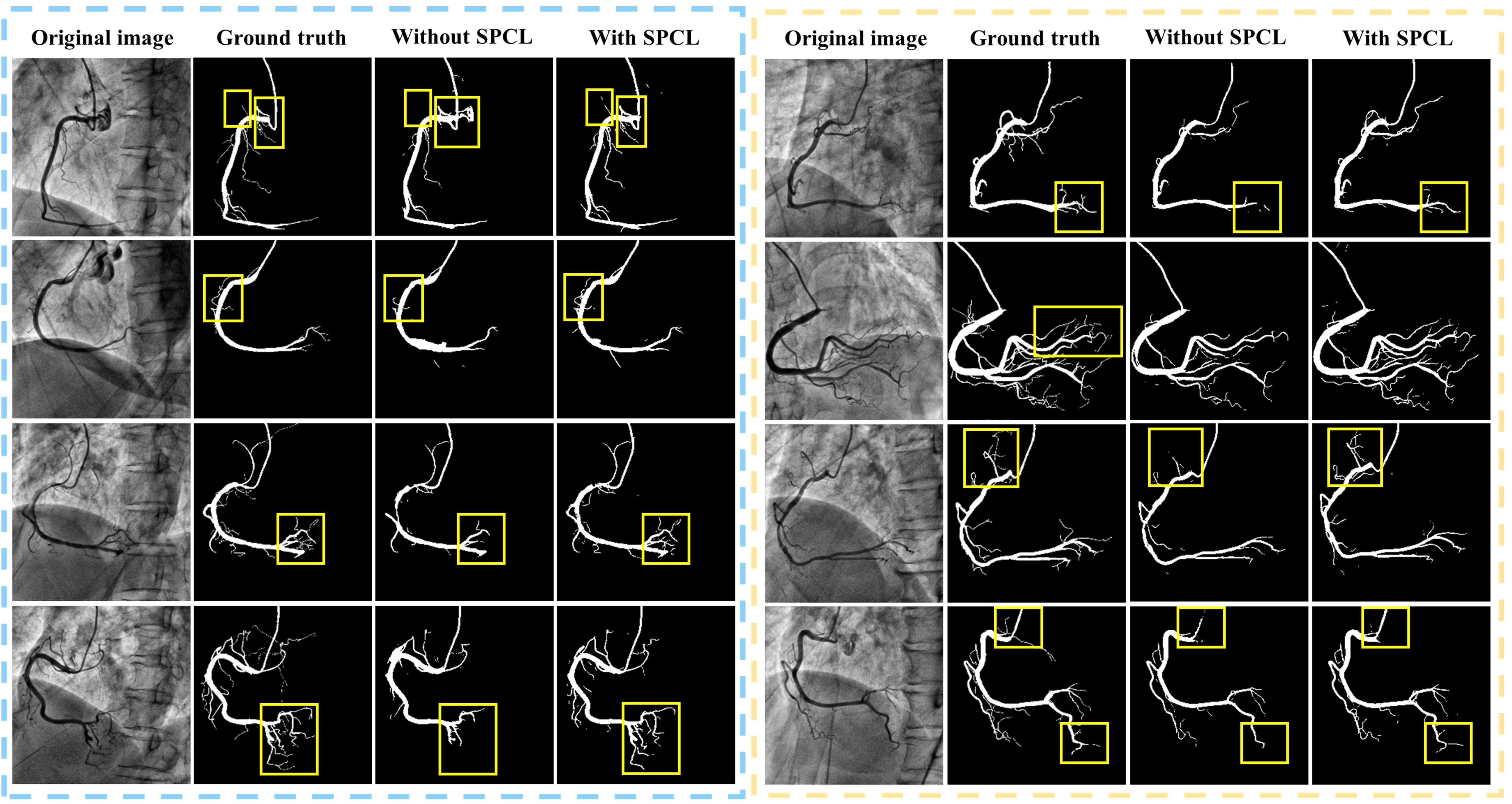}
	\caption{Visualization comparison of R2AU-Net (blue border) and R2U-Net (orange border) segmentation with/without SPCL loss on coronary DSA images. Left to right: original angiogram, ground truth,$\gamma=0$, and SPCL-enhanced results ($\gamma=1$). Yellow rectangles highlight SPCL's improvements in vessel continuity and false positive reduction.}      
	\label{fig10}
        \vspace{-15pt}
\end{figure}

\subsection{Comparison results}
We present our findings through two complementary analyses. First, we evaluate the efficacy of our proposed Supervised Prototypical Contrastive Loss (SPCL) across six established medical image segmentation architectures: U-Net\cite{ronneberger2015u}, Att U-Net\cite{oktay2018attention}, UNet++\cite{zhou2018unet++}, R2U-Net\cite{alom2018recurrent}, Attention UNet++\cite{li2020attention}, and R2AU-Net\cite{zuo2021r2au}. For each architecture, we conduct paired experiments comparing baseline performance (\(\gamma=0\) in Eq.~\ref{eq9}) against SPCL-enhanced versions (\(\gamma=1\) in Eq.~\ref{eq9}). 

As shown in Table~\ref{table1}, our 5-fold cross-validated results on the coronary DSA dataset demonstrate consistent and statistically significant improvements when employing SPCL. Specifically, the Dice coefficient increases from 87.33\%, 87.49\%, 86.81\%, 86.21\%, 87.15\%, and 85.68\% to 87.87\%, 87.85\%, 87.83\%, 87.03\%, 87.79\%, and 86.80\% for U-Net,  Att U-Net, UNet++, R2U-Net, Attention UNet++, and R2AU-Net respectively. Furthermore, our comparative analysis of loss functions for coronary DSA segmentation (Fig.~\ref{fig9}) reveals that integrating SPCL with Dice+BCE yields consistent improvements in all architectures, with DSC gains ranging from +0.36 (U-Net) to +1.12 (R2AU-Net).

\begin{table*}[htbp]
  \centering
  \caption{Performance comparison of state-of-the-art medical image segmentation models on the coronary DSA dataset. Results are presented as (means $\pm$ standard deviations) across 5-fold cross-validation. Statistical significance of performance improvements achieved by \texttt{MSA-UNet3+} is verified using a one-tailed $t$-test.}
    \begin{tabular}{ l l l l l l }
    \toprule
    \multicolumn{1}{l}{\textbf{Architecture }} & \multicolumn{1}{c}{\textbf{Recall $\uparrow$}} & \multicolumn{1}{c}{\textbf{F1 $\uparrow$}} & \multicolumn{1}{c}{\textbf{Dice $\uparrow$}} & \multicolumn{1}{c}{\textbf{ASD $\downarrow$}} & \multicolumn{1}{c}{\textbf{ACD $\downarrow$}} \\
    \midrule
    \multicolumn{1}{l}{Swinunet \cite{cao2022swin}} & \multicolumn{1}{l}{79.14±0.0052$\star$} & \multicolumn{1}{l}{81.95±0.0044$\star$} & \multicolumn{1}{l}{81.82±0.0047$\star$} & \multicolumn{1}{l}{1.77±0.1565$\star$} & \multicolumn{1}{l}{1.74±0.1615$\star$} \\[0.1cm]
     \multicolumn{1}{l}{CMU\_Net \cite{tang2023cmu}} & \multicolumn{1}{l}{85.14±0.0209$\diamond$} & \multicolumn{1}{l}{\underline{87.45±0.0070}$\diamond$$\diamond$} & \multicolumn{1}{l}{\underline{87.40±0.0070}$\diamond$} & \multicolumn{1}{l}{0.85±0.0676$\nabla$} & \multicolumn{1}{l}{0.83±0.0623$\nabla$} \\[0.1cm]
    \multicolumn{1}{l}{MMDC\_Net \cite{zhong2022you}} & \multicolumn{1}{l}{84.56±0.0179$\nabla$} & \multicolumn{1}{l}{87.11±0.0051$\nabla$} & \multicolumn{1}{l}{87.06±0.0051$\nabla$} & \multicolumn{1}{l}{0.90±0.1070$\nabla$} & \multicolumn{1}{l}{0.87±0.0917$\nabla$$\nabla$} \\[0.1cm]
    \multicolumn{1}{l}{FR\_UNet \cite{liu2022full}} & \multicolumn{1}{l}{85.98±0.0076$\diamond$} & \multicolumn{1}{l}{87.28±0.0046$\nabla$} & \multicolumn{1}{l}{87.23±0.0046$\nabla$} & \multicolumn{1}{l}{0.92±0.0422$\star$$\star$} & \multicolumn{1}{l}{0.90±0.0390$\star$} \\[0.1cm]
    \multicolumn{1}{l}{Isunetv1 \cite{pu2023semi}} & \multicolumn{1}{l}{81.67± 0.0055$\star$} & \multicolumn{1}{l}{84.45± 0.0036$\star$} & \multicolumn{1}{l}{84.37± 0.0037$\star$} & \multicolumn{1}{l}{1.37± 0.0755$\star$} & \multicolumn{1}{l}{1.33±0.0757$\star$} \\[0.1cm]
    \multicolumn{1}{l}{CA\_Net \cite{xie2023canet}} & \multicolumn{1}{l}{85.05±0.0143$\nabla$} & \multicolumn{1}{l}{87.23±0.0046$\nabla$} & \multicolumn{1}{l}{87.18±0.0046$\nabla$} & \multicolumn{1}{l}{0.86±0.0515$\nabla$$\nabla$} & \multicolumn{1}{l}{0.83±0.0440$\nabla$$\nabla$} \\[0.1cm]
    \multicolumn{1}{l}{MBS\_Net \cite{jin2023novel}} & \multicolumn{1}{l}{84.88±0.0077$\nabla$$\nabla$} & \multicolumn{1}{l}{86.94±0.0053$\nabla$$\nabla$} & \multicolumn{1}{l}{86.88±0.0053$\nabla$$\nabla$} & \multicolumn{1}{l}{0.90±0.0685$\nabla$$\nabla$} & \multicolumn{1}{l}{0.88±0.0621$\nabla$$\nabla$} \\[0.1cm]
    \multicolumn{1}{l}{CMU\_NeXt \cite{tang2024cmunext}} & \multicolumn{1}{l}{82.57±0.0122$\star$} & \multicolumn{1}{l}{86.66±0.0045$\nabla$$\nabla$} & \multicolumn{1}{l}{86.61±0.0046$\nabla$$\nabla$} & \multicolumn{1}{l}{0.96±0.0562$\star$} & \multicolumn{1}{l}{0.91±0.0473$\star$} \\[0.1cm]
    \multicolumn{1}{l}{BCU\_Net \cite{zhang2023bcu}} & \multicolumn{1}{l}{83.85±0.0073$\star$} & \multicolumn{1}{l}{85.99±0.0038$\diamond$} & \multicolumn{1}{l}{85.92±0.0038$\star$} & \multicolumn{1}{l}{1.07±0.0421$\star$} & \multicolumn{1}{l}{1.03±0.0374$\star$} \\[0.1cm]
    \multicolumn{1}{l}{MCDAU\_Net \cite{zhou2023dual}} & \multicolumn{1}{l}{\textbf{87.71±0.0276}} & \multicolumn{1}{l}{87.15±0.005$\nabla$} & \multicolumn{1}{l}{87.10±0.0050$\nabla$} & \multicolumn{1}{l}{\underline{0.84±0.0869}$\nabla$} & \multicolumn{1}{l}{\underline{0.82±0.0818$\nabla$}} \\[0.1cm]
    \multicolumn{1}{l}{MGA\_Net \cite{gao2023multi}} & \multicolumn{1}{l}{85.06±0.0164$\nabla$} & \multicolumn{1}{l}{87.07±0.0096$\diamond$} & \multicolumn{1}{l}{87.02±0.0096$\diamond$} & \multicolumn{1}{l}{0.86±0.1076$\nabla$} & \multicolumn{1}{l}{0.84±0.1073$\nabla$} \\[0.1cm]
    \multicolumn{1}{l}{IMFF\_Net \cite{liu2024imff}} & \multicolumn{1}{l}{84.36±0.0103$\nabla$$\nabla$} & \multicolumn{1}{l}{87.20±0.0073$\diamond$} & \multicolumn{1}{l}{87.15±0.0073$\diamond$} & \multicolumn{1}{l}{0.88±0.0970$\nabla$} & \multicolumn{1}{l}{0.85±0.0945$\nabla$} \\[0.1cm]
    \multicolumn{1}{l}{DATrans\_UNet \cite{sun2024transunet}} & \multicolumn{1}{l}{82.62±0.0089$\star$} & \multicolumn{1}{l}{85.92±0.0069$\star$} & \multicolumn{1}{l}{85.87±0.0071$\star$} & \multicolumn{1}{l}{1.08±0.1151$\star$} & \multicolumn{1}{l}{1.04±0.1009$\star$} \\[0.1cm]
    \multicolumn{1}{l}{PMFS\_Net \cite{zhong2025pmfsnet}} & \multicolumn{1}{l}{82.85±0.0090$\star$} & \multicolumn{1}{l}{84.55±0.0038$\star$} & \multicolumn{1}{l}{84.49±0.0039$\star$} & \multicolumn{1}{l}{1.33±0.0847$\star$} & \multicolumn{1}{l}{1.26±0.0673$\star$} \\[0.1cm]
    \textbf{MSA-UNet3+ (ours)} & \underline{86.78±0.008} & \textbf{87.78±0.0036} & \textbf{87.73±0.0036} & \textbf{0.76±0.0414} & \textbf{0.74±0.0367} \\[0.1cm]
    \bottomrule
    \end{tabular}%
  \label{table2}%
  \vspace{0.10cm} 
  \begin{minipage}{\textwidth}
    \small 
    \textbf{Notes:} 
    Bold numbers indicate the best performance for each architecture and metric, while underlined numbers denote the second-best performance. $\star$ \(p-value < 0.0001\), $\star$$\star$ \(p-value < 0.0005\), $\nabla$ \(p-value < 0.05\), $\nabla$$\nabla$ \(p-value < 0.01\), $\diamond$ \(p-value < 0.1\), $\diamond$$\diamond$ \(p-value < 0.2\) compared with MSA-UNet3+(ours).
  \end{minipage}
  \vspace{-12pt}
\end{table*}

\begin{figure}
	\centering
	\includegraphics[width=0.7\columnwidth]{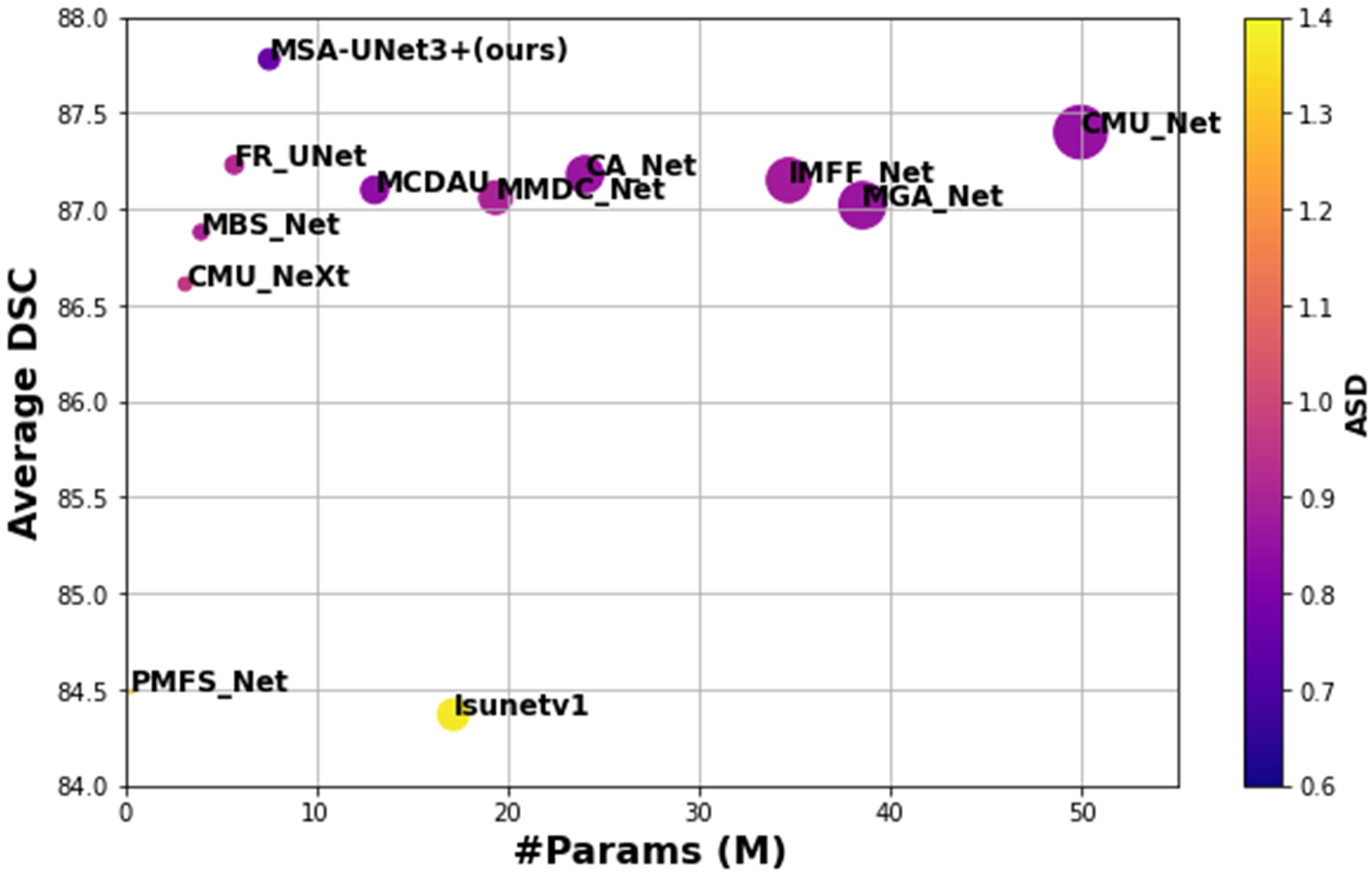}
	\caption{Comparison of Deep Learning Models for Coronary DSA Image Segmentation: Model Efficiency vs. Segmentation Accuracy. The figure illustrates the trade-off between model efficiency (parameters in millions) and segmentation accuracy (average DSC and ASD) for various deep learning models, including the proposed MSA-UNet3+. The size and color of each scatter point represent the model size and ASD values, respectively, with the color bar providing a visual mapping.}
	\label{fig7}
\end{figure}

\begin{figure}
	\centering
	\includegraphics[width=0.7\columnwidth]{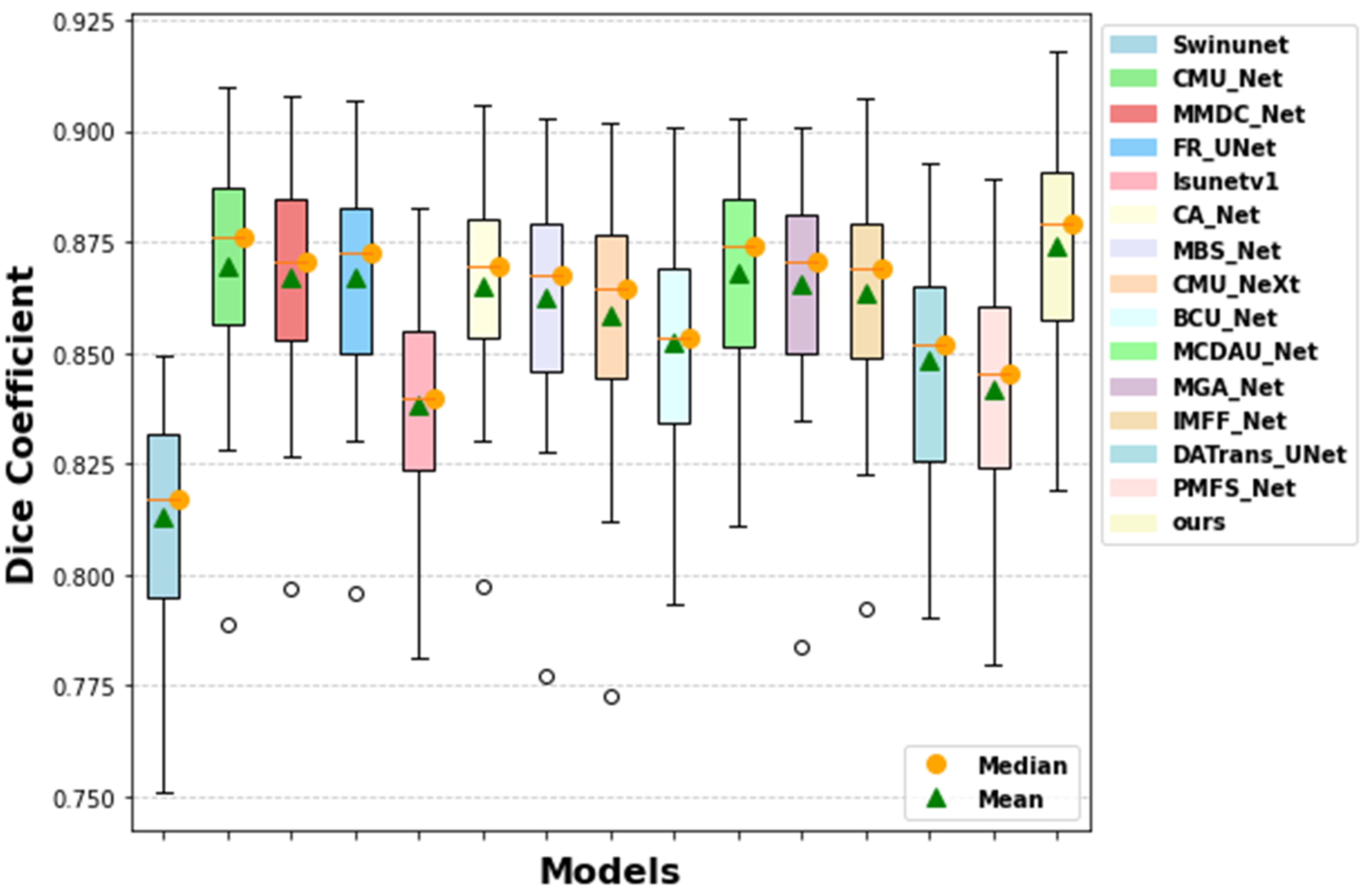}
	\caption{A Boxplot of the Dice coefficient is shown for all test samples of the coronary DSA dataset. In the boxes, orange line and green triangle within each box represents the median and the mean, respectively.}
	\label{fig8}
\end{figure}

Qualitatively, Fig.~\ref{fig10} demonstrates the effectiveness of SPCL through two key improvements: (1) enhanced vessel continuity, particularly in complex branching structures where the baseline model fragments vessels, and (2) superior preservation of thin vessels that conventional approaches often miss. 

Complementing our SPCL evaluation, the second perspective evaluates the proposed MSA-UNet3+ with SPCL loss against fifteen state-of-the-art medical segmentation models: SwinUNet \cite{cao2022swin}, CMU\_Net \cite{tang2023cmu}, MMDC\_Net \cite{zhong2022you}, FR\_Unet \cite{liu2022full}, Isunetv1 \cite{pu2023semi}, CA\_Net \cite{xie2023canet}, MBS\_Net \cite{jin2023novel}, CMU\_NeXt \cite{tang2024cmunext}, BCU\_Net \cite{zhang2023bcu}, MCDAU\_Net \cite{zhou2023dual}, MGA\_Net \cite{gao2023multi}, IMFF\_Net \cite{liu2024imff}, DATrans\_Unet \cite{sun2024transunet}, and PMFS\_Net \cite{zhong2025pmfsnet}. 

\subsubsection{Quantitative analysis}
This section assesses the segmentation performance of MSA-UNet3+ with SPCL loss on the coronary DSA  dataset. The quantitative results in Table~\ref{table2} demonstrate the superior performance of the model, particularly in capturing fine vascular structures compared to existing methods.  

Fig.~\ref{fig7} illustrates the trade-off between model efficiency (measured by the number of parameters) and segmentation accuracy (quantified by DSC and ASD). The proposed
MSA-UNet3+ achieves an optimal balance, attaining a DSC
of 87.78\% with only 7.54 million parameters---significantly fewer than comparably accurate models such as CA\_Net, CMU\_Net, and MMDC\_Net. While PMFS\_Net demonstrates greater efficiency (0.33M parameters), its lower DSC (84.49\%) highlights the accuracy-efficiency compromise. Visual encoding in the figure enhances interpretation: the point size scales with model size, while the color intensity reflects ASD, with darker colors indicating lower ASD (better surface distance accuracy). MSA-UNet3+ excels on both axes, combining competitive ASD (0.76) with compact architecture. 

Further validating the model's robustness, the performance distribution analysis in Fig.~\ref{fig8} demonstrates that our approach (\textit{``ours''}) achieves both a high median Dice coefficient and a relatively tight interquartile range, suggesting reliable segmentation performance. The visualization's annotated markers (orange circles for medians, green triangles for means) reveal an important trade-off: while models like \textit{CMU\_Net} and \textit{MMDC\_Net} show higher median Dice coefficients, they have wider interquartile ranges, indicating less consistency in performance. In contrast, more stable models (\textit{MGA\_Net}, \textit{MCDAU\_Net}) achieve narrower ranges but with reduced segmentation quality. 
\begin{figure}
	\centering
	\includegraphics[width=1.0\columnwidth]{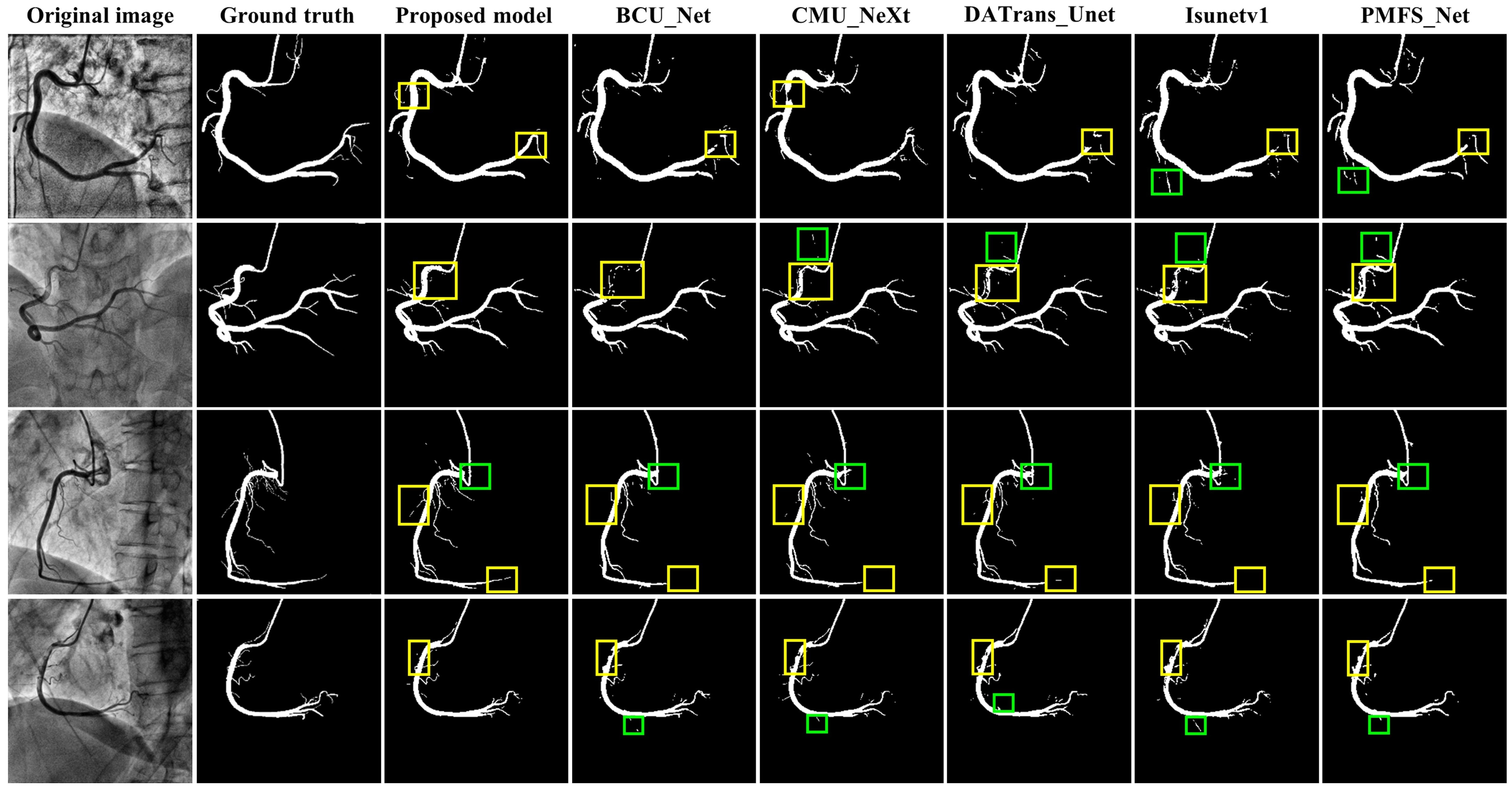}
	\caption{Qualitative Results: The four rows show the four test samples, and the eight columns show the DSA images, which are (from left to right): original image, ground truth, proposed model results, BCU\_Net, CMU\_NeXt, DATrans\_Unet, Isunetv1, and PMFS\_Net. Yellow rectangles highlight false negatives, and green rectangles indicate false positives.}
	\label{fig6}
        \vspace{-15pt}
\end{figure}
\subsubsection{Qualitative analysis}
The qualitative comparison presented in Fig.~\ref{fig6} examines four representative test samples, each displaying from left to right: the original DSA image, ground truth annotation, segmentation outputs from five established methods (BCU\_Net, CMU\_NeXt, DATrans\_Unet, Isunetv1, and PMFS\_Net), and the results from our proposed approach. Visual annotations highlight performance differences through yellow rectangles marking false negatives (missed vessels) and green rectangles indicating false positives (spurious detections). 
\section{Discussion}
\label{sec:discussion}
\subsection{Effectiveness of the supervised prototypical contrastive loss}
The quantitative improvements across all architectures and metrics confirm SPCL's effectiveness as a generalizable enhancement for medical image segmentation tasks. The performance progression aligns with the findings presented in Table~\ref{table1} and Fig.~\ref{fig9}, confirming the dual ability of SPCL to: (1) improve feature discrimination through semantic clustering, and (2) address class imbalance through contrastive learning. The demonstrated robustness across architectures (from U-Net to R2AU-Net) and the magnitude of improvement position SPCL as a general solution for medical segmentation tasks.

These results corroborate our earlier quantitative assessments while providing new insights into SPCL's architecture-dependent optimization characteristics. The visual improvements shown in Fig.~\ref{fig10} align directly with our quantitative metrics, illustrating how the measured gains in Dice coefficient translate to clinically relevant outcomes such as enhanced vessel continuity and preservation of thin structures, the yellow-highlighted regions specifically showcase SPCL's ability to maintain anatomical fidelity in clinically challenging areas, preserving critical vessel connectivity while suppressing background noise. The consistent performance gains across diverse model architectures suggest that SPCL operates effectively at the feature representation level, complementing rather than competing with architectural innovations. This is further evidenced by the superior performance of MSA-UNet3+ with SPCL when compared against fifteen state-of-the-art models. The combined approach demonstrates particular efficacy in challenging anatomical regions, likely due to SPCL's ability to enforce better feature separation and MSA-UNet3+'s capacity for multi-scale feature integration.

The findings suggest that integrating contrastive learning principles through SPCL provides a viable pathway for improving segmentation fidelity in complex medical imaging tasks, offering benefits that are complementary to architectural advancements.
\begin{table*}[htbp]
  \centering
  \caption{Ablation study of the proposed MSA-UNet3+ framework, presenting performance metrics for eight configurations to assess the contributions of the MSD-Bottleneck, CAFM, and SPCL modules. Symbols denote (\ding{51}) presence and (\ding{55}) absence of each component in the ablation study.}
  \label{tab:addlabel}
  \setlength{\tabcolsep}{7pt}
  \renewcommand{\arraystretch}{1.1}
  \begin{tabular}{@{}c@{\hspace{4pt}}ccccccccc@{}}
    \toprule
    \multicolumn{4}{c}{\textbf{Configuration}} & & \multicolumn{5}{c}{\textbf{Metric}} \\
    \cmidrule(lr){1-4} \cmidrule(l){6-10} 
    \textbf{Idx} & \textbf{SPCL} & \textbf{CAFM} & \textbf{MSD} & & \textbf{Recall $\uparrow$} & \textbf{F1 $\uparrow$} & \textbf{Dice $\uparrow$} & \textbf{ASD $\downarrow$} & \textbf{ACD $\downarrow$} \\ [0.1cm]
    \midrule
    1 & \ding{55} & \ding{55} & \ding{55} & & 85.21±0.0057 & 87.23±0.0037 & 87.18±0.0038 & 0.8914±0.0465 & 0.8639±0.0452 \\ [0.1cm]
    2 & \ding{51} & \ding{55} & \ding{55} & & 85.31±0.0051 & 87.42±0.0045 & 87.37±0.0046 & 0.8517±0.0681 & 0.8263±0.0609 \\ [0.1cm]
    3 & \ding{55} & \ding{51} & \ding{55} & & 85.99±0.0107 & 87.59±0.0040 & 87.55±0.0041 & 0.8003±0.0464 & 0.7820±0.0413 \\ [0.1cm]
    4 & \ding{51} & \ding{51} & \ding{55} & & 85.83±0.0043 & 87.63±0.0028 & 87.59±0.0028 & 0.7923±0.0519 & 0.7722±0.0460 \\ [0.1cm]
    5 & \ding{55} & \ding{55} & \ding{51} & & 86.53±0.0088 & 87.58±0.0034 & 87.54±0.0033 & 0.8136±0.0656 & 0.7989±0.0614 \\ [0.1cm]
    6 & \ding{51} & \ding{55} & \ding{51} & & 86.31±0.0065 & 87.73±0.0026 & 87.68±0.0025 & 0.7689±0.0395 & 0.7509±0.0372 \\ [0.1cm]
    7 & \ding{55} & \ding{51} & \ding{51} & & 85.95±0.0065 & 87.62±0.0045 & 87.58±0.0045 & 0.8098±0.0577 & 0.7915±0.0578 \\ [0.1cm]
    8 & \ding{51} & \ding{51} & \ding{51} & & \textbf{86.78±0.0080} & \textbf{87.78±0.0036} & \textbf{87.73±0.0036} & \textbf{0.7587±0.0414} & \textbf{0.7411±0.0367} \\
    \bottomrule
  \end{tabular}%
  \label{table3}%
\end{table*}%
\subsection{Performance of MSA-UNet3+ against state-of-the-art models}
Through comprehensive quantitative metrics and qualitative visual comparisons, we demonstrate MSA-UNet3+'s superior segmentation performance, particularly in challenging anatomical regions. The quantitative results in Table~\ref{table2} demonstrate the superior performance of the model, particularly in capturing fine vascular structures compared to existing methods.

Evaluation of our proposed MSA-UNet3+ on the coronary DSA dataset (Table~\ref{table2}) reveals competitive performance across key metrics. Although MCDAU\_Net attains the highest recall ($87.71 \pm 0.0276$), MSA-UNet3+ achieves superior performance in several critical aspects: the second-highest recall ($86.78 \pm 0.0080$), the best F1 score ($87.78 \pm 0.0036$) and Dice coefficient ($87.73 \pm 0.0036$) among all models, and exceptional boundary precision with the lowest ASD ($0.76 \pm 0.0414$) and ACD ($0.74 \pm 0.0367$). While CMU\_Net (recall: $85.14 \pm 0.0209$) and MCDAU\_Net demonstrate competitive results, they underperform MSA-UNet3+ in overall F1, Dice, and boundary accuracy. Similarly, FR\_UNet and CA\_Net show competent but consistently inferior performance relative to our proposed model.

According to Fig.~\ref{fig6} Our method exhibits consistently superior performance across all evaluated samples, manifesting three distinct advantages: (1) precise delineation of fine vascular structures, particularly visible in row 3's thin vessels; (2) reliable performance in diagnostically challenging areas containing imaging noise and anatomical overlaps; and (3) balanced mitigation of both false positive and false negative errors relative to comparator methods. Detailed analysis reveals that while BCU\_Net and CMU\_NeXt produce substantial false positive detections in complex anatomical backgrounds, DATrans\_Unet and Isunetv1 exhibit pronounced false negative rates, failing to identify fine vascular details. PMFS\_Net demonstrates intermediate performance with variable accuracy across different vascular structures.

These quantitative and qualitative results establish MSA-UNet3+ as a state-of-the-art solution for coronary DSA segmentation, combining a balanced metric performance with clinically crucial boundary precision. The close visual correspondence between our segmentation results and the ground truth annotations highlights the model's potential for clinical implementation in coronary artery analysis, where precise vessel delineation is critical for accurate diagnosis and treatment planning.

\subsection{Ablation study}
The proposed MSA-UNet3+ framework integrates three key innovations to address the challenges of coronary DSA segmentation: the Multi-scale Dilated Bottleneck (MSD-Bottleneck) for the extraction of multi-resolution features, the Contextual Attention Fusion Module (CAFM) for spatial-contextual information integration, and the Supervised Prototypical Contrastive Loss (SPCL) for feature discrimination and class imbalance mitigation. Together, these components target complex vessel morphology, foreground-background imbalance, and boundary ambiguity in coronary angiography. As demonstrated in Table~\ref{table3}, an ablation study quantifies the contribution of each module to the overall performance of the model, highlighting their complementary strengths in addressing distinct segmentation challenges.
\subsubsection{Effect of the supervised prototypical contrastive loss (SPCL)}
The SPCL represents a specialized loss function engineered to improve feature discrimination by promoting more distinct class representations within the learned feature space. When incorporated into the baseline model, SPCL generates consistent improvements across all evaluation metrics: Recall exhibits an increase from 85.212\% to 85.305\%, while both the F1-score and Dice coefficient show measurable gains, improving from 87.233\% and 87.181\% to 87.418\% and 87.367\%, respectively. Most notably, SPCL produces substantial reductions in segmentation error met-
rics, 
with the Average Surface Distance (ASD) decreasing from 0.8914 to 0.8517 and the Average Contour Distance (ACD) improving from 0.8639 to 0.8263. These results demonstrate SPCL’s ability to sharpen class separation, by refining the model’s feature representations. The complete potential of SPCL becomes particularly apparent when integrated with complementary architectural components. For example, when combined with the Multi-Scale Dilated Bottleneck (baseline+MSD+SPCL configuration), SPCL enables significant performance enhancements. The F1-score and Dice coefficient rise to 87.726\% and 87.680\%, respectively, while the ASD and ACD metrics demonstrate further improvement, reaching 0.7689 and 0.7509. These results represent a marked advancement over the baseline performance. This combination highlights SPCL’s ability to complement multi-scale feature extraction, enabling the model to distinguish between classes with remarkable performance across varying scales. 
\subsubsection{Impact of the multi-scale dilated bottleneck (MSD-bottleneck)}
The MSD-Bottleneck module employs dilated convolutions with varying dilation rates to capture multi-scale features, thereby enabling robust handling of objects with diverse sizes and morphological characteristics, a critical capability for coronary DSA image segmentation. When integrated into the baseline architecture, the MSD-Bottleneck produces substantial performance gains, increasing recall from 85.212\% to 86.530\% while simultaneously reducing both the average surface distance (ASD) from 0.8914 to 0.7989 and the average contour distance (ACD) from 0.8639 to 0.8136. These quantitative improvements underscore the fundamental importance of multi-scale feature extraction for accurate segmentation of vascular structures across different spatial scales. 

The module's effectiveness is further amplified when combined with the Supervised Prototypical Contrastive Loss (SPCL), as evidenced by additional reductions in segmentation error metrics to 0.7689 (ASD) and 0.7509 (ACD). This complementary interaction between MSD-Bottleneck and SPCL arises from their complementary mechanisms: while MSD-Bottleneck provides comprehensive multi-scale feature extraction through its dilated convolution architecture, SPCL concurrently optimizes the discriminative quality of these features. The combined operation of these components yields superior segmentation performance compared to their individual contributions, demonstrating that multi-scale feature capture and feature space refinement operate in a mutually reinforcing manner.
\subsubsection{Role of the contextual attention fusion module (CAFM)}
The CAFM Module enhances segmentation performance by selectively integrating contextual information from diverse image regions while suppressing irrelevant features. This attention mechanism substantially improves the model's global context comprehension, a critical factor for precise coronary vessel segmentation. When implemented with the baseline architecture (\texttt{baseline+CAFM}), the module demonstrates measurable performance improvements: Recall increases from 85.212\% to 85.990\%, accompanied by rises in both F1-score (87.233\% to 87.593\%) and Dice coefficient (87.181\% to 87.547\%). Furthermore, CAFM reduces segmentation errors substantially, as evidenced by decreases in Average Surface Distance (ASD) from 0.8914 to 0.8003 and Average Contour Distance (ACD) from 0.8639 to 0.7820, confirming enhanced boundary alignment precision. However, the module's incremental benefits become less substantial when combined with other advanced components. 

In configurations combining CAFM with either the Supervised Prototypical Contrastive Loss (\texttt{baseline+SPCL+CAFM}) or the Multi-Scale Dilated Bottleneck (\texttt{baseline+CAFM+MSD}), performance gains are more modest, showing only slight improvements in F1-scores and minimal reductions in ASD and ACD metrics. This observation suggests that while CAFM independently provides significant contextual understanding enhancements, its additive value diminishes when integrated with other sophisticated modules, as their respective contributions appear to operate through complementary rather than amplifying mechanisms with respect to CAFM's functionality.
\subsubsection{Combined contributions}
The complete MSA-UNet3+ with SPCL loss architecture, integrating all three key components (MSD-Bottleneck, CAFM, and SPCL), demonstrates optimal performance across all evaluation metrics. The unified model achieves a Recall of 86.775\%, along with an F1-score of 87.776\% and Dice coefficient of 87.733\%, while simultaneously attaining the most favorable boundary alignment metrics with ASD and ACD values of 0.7587 and 0.7411 respectively. These results illustrate the effective collaborative operation of the constituent modules: the MSD-Bottleneck's multi-scale feature extraction capability, CAFM's contextual information integration, and SPCL's enhanced feature discrimination collectively establish a comprehensive and precise segmentation framework. Each component addresses specific challenges in coronary vessel segmentation, and their combined operation enables superior performance in handling complex anatomical variations. Through systematic ablation analysis, we confirm that every module (MSD-Bottleneck, CAFM, and SPCL) makes substantial and distinct contributions to the overall performance of MSA-UNet3+. The complete integration of these components yields segmentation results that surpass current state-of-the-art approaches, positioning MSA-UNet3+ as a particularly effective solution for the demanding requirements of coronary DSA image analysis. 

\subsubsection{Potential adaptation to the left coronary artery and other vessels}
While the current study was validated on a right coronary artery dataset, the proposed MSA-UNet3+ framework and the Supervised Prototypical Contrastive Loss (SPCL) are inherently designed to handle diverse vascular morphologies and imaging variations. The left coronary artery presents distinct anatomical characteristics---such as broader bifurcation patterns, larger lumen diameters, and often more complex overlap with surrounding structures. These challenges are directly addressed by our model's innovations: the MSD-Bottleneck captures multi-scale bifurcation patterns; the CAFM integrates contextual cues to resolve overlaps; and the SPCL mitigates class imbalance caused by complex backgrounds---making our framework particularly suitable for left coronary artery segmentation without architectural modifications. Adaptation would primarily involve fine-tuning on a left coronary artery dataset, leveraging our pre-trained weights as a strong initialization.

Furthermore, the framework's modular design and general-purpose contrastive learning strategy allow for straightforward adaptation to other vascular domains (e.g., cerebral, retinal, or peripheral vessels) where similar challenges of fine structures, complex backgrounds, and class imbalance exist. For instance, in retinal vessel segmentation, the need to distinguish fine capillaries from a heterogeneous background aligns well with our approach. Successful extension to these domains would further validate the core contribution of our work: a generalizable segmentation framework that addresses fundamental challenges in medical imaging through multi-scale attention and contrastive feature learning.

Future work will include systematic validation on multi-vessel coronary datasets and extension to other angiographic modalities, leveraging the same principles to address varied vessel geometries and imaging conditions.

\vspace{18pt}
\section{Conclusion}
\label{sec:conclusion}
In this study, we introduced the MSA-UNet3+, a novel architecture designed to enhance coronary Digital Subtraction Angiography (DSA) image segmentation. The framework integrates a multi-scale dilated bottleneck (MSD-bottleneck) with contextual attention fusion modules (CAFM) to enable precise multi-scale feature extraction while preserving fine vessel details—critical for capturing complex vascular structures. Our technique is centered on the Supervised Prototypical Contrastive Loss (SPCL), which significantly improves the model’s ability to handle class imbalance and high intra-class variance in coronary DSA images. By integrating supervised contrastive learning with prototypical contrastive learning, SPCL effectively clusters similar features while distinguishing challenging background samples. This dual mechanism not only enhances the semantic embeddings of the encoder but also ensures that the model focuses on learning the most informative features for segmentation, particularly foreground and hard-to-classify background samples, complementing the architecture's multi-scale capabilities. Our experimental results demonstrate that MSA-UNet3+ outperforms existing state-of-the-art methods, achieving a Dice coefficient of 87.73\% and an F1-score of 87.78\%. The architecture's ability to maintain robust performance in challenging imaging conditions underscores its clinical relevance in diagnosing coronary artery diseases.

However, the current framework faces limitations in severe noise, motion artifacts, or extremely low contrast, where vascular structures become indistinct. These challenging scenarios currently lead to suboptimal segmentation outcomes due to the inherent limitations of angiographic imaging. Future research directions will focus on three key improvements: First, we will investigate multimodal fusion approaches incorporating complementary imaging modalities such as intravascular ultrasound (IVUS) and optical coherence tomography (OCT) to enhance segmentation reliability in problematic cases. Second, we plan to develop optimized lightweight variants of the architecture suitable for deployment on medical imaging devices, thereby facilitating real-time clinical decision-making during interventional procedures. Third, although the current implementation specifically targets coronary DSA analysis, the underlying methodology shows considerable potential for adaptation to other medical segmentation tasks characterized by similar challenges, including class imbalance and complex anatomical backgrounds. 
\\
\\
\textbf{CRediT authorship contribution statement}
\\
\\
\textbf{Rayan Merghani Ahmed}: Conceptualization, Methodology, Software , Visualization, Writing – original draft. 
\textbf{Adnan Iltaf}: Writing – review \& editing. \textbf{Mohamed Elmanna}: Validation, Writing – review \& editing \textbf{Gang Zhao}: Data Curation, labeling. \textbf{Hongliang Li}: Data Curation, labeling. \textbf{Du Yue}: Project Administration
\textbf{Bin Li}: Writing – review \& editing, Supervision. \textbf{Shoujun Zhou}: Writing – review \& editing, Supervision, Funding acquisition.
\\
\\
\textbf{Declaration of competing interest}
\\
\\
The authors declare that they have no conflicts of interest.
\\
\\
\textbf{Acknowledgements}
\\
\\
This work was supported by the Shenzhen Medical Research Fund [No. D2404001]; and in part by the Key Research and Development Program of Guangdong Province [No. 2025B1111020001]; the Shenzhen Municipal STIB Key programs [No. CJGJZD20230724093303007, and KJZD20240903101259001]; National Key Laboratory of the CAS on Medical Imaging Science and Technology System, the Xisike Clinical Oncology Research Foundation[Y-2024AZ(NSCLC)MS-0156]; and SIAT-WUXI Joint Innov-Group for AGI-MET.
\\
\\
\textbf{Data availability}
\\
\\
The authors do not have permission to share data.

\bibliographystyle{unsrtnat}
\bibliography{cas-refs_u}

\end{document}